\documentclass[letterpaper,10pt]{JHEP3}

\usepackage{amsfonts,amssymb,amsmath,amsxtra,graphicx,revsymb}

\def\circa#1{\,\raise.3ex\hbox{$#1$\kern-.75em\lower1ex\hbox{$\sim$}}\,}
\makeatletter

\renewcommand{\theequation}{\thesection.\arabic{equation}}

\newcommand{\mpl}{M_{\rm Pl}}
\newcommand{\csch}{{\rm csch}}
\newcommand{\sech}{{\rm sech}}
\newcommand{\calA}{{\cal A}}
\newcommand{\calB}{{\cal B}}
\newcommand{\calC}{{\cal C}}
\newcommand{\calD}{{\cal D}}
\newcommand{\calF}{{\cal F}}
\newcommand{\calG}{{\cal G}}

\newcommand{\calK}{{\cal K}}

\newcommand{\calP}{{\cal P}}

\newcommand{\calR}{{\cal R}}

\def\circa#1{\,\raise.3ex\hbox{$#1$\kern-.75em\lower1ex\hbox{$\sim$}}\,}
\makeatletter

\def\art{\@ifnextchar[{\eart}{\oart}}
\def\eart[#1]#2#3#4#5#6{{\rm #2}, {\em #3  #4} {\rm (#6) #5} ({\em #1})}
\def\hepart[#1]#2{{\rm #2, \em#1}}
\newcommand{\oart}[5]{{\rm #1}, {\em #2  #3} {\rm (#5) #4}}

\newcounter{alphaequation}[equation]
\def\thealphaequation{\theequation\hbox to
0.6em{\hfil\alph{alphaequation}\hfil}}
\def\eqnsystem#1{
\def\@eqnnum{{\rm (\thealphaequation)}}
\def\@@eqncr{\let\@tempa\relax \ifcase\@eqcnt \def\@tempa{& & &} \or
  \def\@tempa{& &}\or \def\@tempa{&}\fi\@tempa
  \if@eqnsw\@eqnnum\refstepcounter{alphaequation}\fi
\global\@eqnswtrue\global\@eqcnt=0\cr}
\refstepcounter{equation} \let\@currentlabel\theequation \def\@tempb{#1}
\ifx\@tempb\empty\else\label{#1}\fi
\refstepcounter{alphaequation}
\let\@currentlabel\thealphaequation
\global\@eqnswtrue\global\@eqcnt=0 \tabskip\@centering\let\\=\@eqncr
$$\halign to \displaywidth\bgroup \@eqnsel\hskip\@centering
$\displaystyle\tabskip\z@{##}$&\global\@eqcnt\@ne
\hskip2\arraycolsep\hfil${##}$\hfil& \global\@eqcnt\tw@\hskip2\arraycolsep
$\displaystyle\tabskip\z@{##}$\hfil
\tabskip\@centering&\llap{##}\tabskip\z@\cr}
\def\endeqnsystem{\@@eqncr\egroup$$\global\@ignoretrue} \makeatother

\title{Towards multi-field D-brane inflation in a warped throat}

\author{Heng-Yu Chen$^a$, Jinn-Ouk Gong$^b$, Kazuya Koyama$^c$ and Gianmassimo Tasinato$^d$
\\
$^a$ \textit{Department of Physics, University of Wisconsin-Madison, Madison, WI 53706, USA}
\\
$^b$ \textit{Instituut-Lorentz for Theoretical Physics, Universiteit Leiden, 2333 CA Leiden, The Netherlands}
\\
$^c$ \textit{Institute of Cosmology and Gravitation, University of Portsmouth, Portsmouth PO1 3FX, UK}
\\
$^d$ \textit{Institut f\"ur Theoretische Physik, Universit\"at Heidelberg, 69120 Heidelberg, Germany}
}

\abstract{
We study the inflationary dynamics in a model of slow-roll inflation in warped throat. Inflation is realized by the motion of a D-brane along the radial direction of the throat, and at later stages instabilities develop in the angular directions. We closely investigate both the single field potential relevant for the slow-roll phase, and the full multi-field one including the angular modes which becomes important at later stages. We study the main features of the  instability process, discussing its possible consequences and identifying the vacua towards which the angular modes are driven.
}

\begin{document}

\section{Introduction and summary}
\label{sec:intro}

Inflation is our most widely accepted paradigm of the very early universe that enables us to explain the observed properties of the cosmic microwave background anisotropies. Present data are well described by the simplest inflationary model, consisting of a single, slowly rolling light scalar field~\cite{Komatsu:2010fb}. However, data are also consistent with more complex models, for example characterized by non-canonical kinetic terms, or involving the dynamics of multiple scalar fields during inflation. Future observations are expected to enable us to distinguish among the various scenarios, provided that we can compute distinctive observational consequences of different setups.

From the theoretical point of view, building models of inflation provides both opportunities and challenges.  Inflation gives a unique opportunity to probe high energy physics since, in its most natural realizations, the characteristic energy scale is well beyond the current or planned particle accelerators. Also, inflation is highly sensitive to its ultraviolet completion: higher dimensional operators contributing to the inflationary potential play a crucial role in determining inflationary dynamics. This implies that inflation is able to probe properties of the theory that underlies a given model. On the other hand, this also means that a detailed knowledge of the setup under consideration is necessary, when embedding models of inflation in fundamental theories like string theory (or its supergravity limit). Usually, one consequence of this ultraviolet sensitivity is that a successful inflationary setup requires careful tunings of the available parameters to avoid the corrections due to higher dimensional operators, which can spoil the  delicate inflationary dynamics and lead to the so-called $\eta$ problem~\cite{Copeland:1994vg,Easson:2009kk}.

Despite these challenges, there have been several attempts to embed consistent models of inflation within string theory: see Ref.~\cite{reprsSI} for representative examples, and Ref.~\cite{SIrevs} for recent reviews. The abundance of moduli fields in the string compactifications, both closed and open, in principle provides us a wide range of possible inflaton candidates. On the other hand, precisely due to the fact that many moduli are usually involved in a given string model, there is the danger that light moduli would interfere with the inflationary process. This is due to the fact that light moduli can gravitationally couple with the inflaton candidate(s), generically leading to the aforementioned $\eta$ problem. Fortunately, over the past ten years new methods have been provided to stabilize undesired moduli within string theory, by means of fluxes and non-perturbative effects, starting with the seminal work of Giddings, Kachru and Polchinski~\cite{Giddings:2001yu}. However, explicitly calculable setups are generally scarce where various required ingredients such as the background geometric fluxes or non-perturbative superpotential\footnote{We however also note that, there have been interesting proposals to geometrize the non-perturbative superpotential in the four dimensional effective field theory, i.e. to replace them by explicit ten-dimensional geometric fluxes~\cite{Marchesano:2009rz,Baumann:2010sx}.} are known.

Given these reasons, the warped deformed conifold~\cite{Klebanov:2000hb} holds a rather special place as an ideal playground for string inflationary model building, leading to the framework of brane inflation in warped throats. To be specific, following the original work~\cite{Kachru:2003sx}, one considers mobile D3-brane(s) moving in such a background geometry, that are attracted by an anti D3-brane located at the tip of the conifold. The setup is sufficiently well understood that the metric, the background geometric fluxes and the moduli stabilization effects are known in detail, and the potential governing the entire inflationary trajectory can be constructed explicitly. Moreover, the parameter space of such a model is also rich enough that semi-realistic inflationary trajectories can be found and compared with observational data. Finally, D3-brane inflation in warped deformed conifold is a rare example among string inflationary models which has a holographic dual description~\cite{Baumann:2010sx,Klebanov:2000hb,Baumann:2008kq}. This offers us a new perspective on the various contributions to the D3-brane motion from dual field theory, and allows us to employ the powerful and highly developed computational techniques of gauge/string duality.

It has been shown that, by suitably tuning the ingredients responsible for stabilizing moduli, it is possible to find examples of inflection point inflation in this scenario~\cite{Cline:2005ty,Baumann:2007ah}. Inflation occurs in regions of the warped throat in which different competing forces act on the moving D3-brane, compensating each other in such a way that the resulting potential is sufficiently flat around an inflection point (see Refs.~\cite{development,Chen:2008ada,Chen:2008ai} for subsequent developments along these lines). In the original model, a Kuperstein-embedded D7-brane on the warped conifold~\cite{Kuperstein:2004hy} was considered, and the non-perturbative gaugino condensation on its worldvolume contributes to stabilize all the moduli besides the inflaton. A sufficient number of $e$-foldings of single field, slow-roll inflation can then be obtained, which can be  geometrically interpreted as a D3-brane moving along the radial direction of the warped throat. All the moduli that do not take part in the inflationary dynamics are made sufficiently massive by fluxes and non-perturbative effects.

In this work, we revisit such an explicit framework, but instead focus on the so-called the Ouyang embedding~\cite{Ouyang:2003df,Chen:2008jj}. The main motivation here is to enrich the properties of the inflationary dynamics, in particular obtaining a framework in which more than one field play a role in the inflationary process. Indeed, in Ref.~\cite{Baumann:2007ah} (see also Refs.~\cite{Burgess:2006cb,Krause:2007jk}), it was anticipated that, for this embedding, moduli fields associated with the D3-brane motion along the angular directions of the warped manifold  become tachyonic towards the tip of the throat. This property is very interesting for realizing a scenario in which the scalar fields associated with angular directions take part in the inflationary dynamics. Indeed, it is known that the evolution of the curvature perturbation after horizon exit, in multiple field scenarios, can have important consequences for the spectrum of the curvature perturbation and non-Gaussianity. The dynamics of light angular fields in brane inflation, in particular in the context of DBI models with non-canonical kinetic terms~\cite{Silverstein:2003hf,Alishahiha:2004eh}, have been widely studied over the past years, starting from Ref.~\cite{anglularinf}, mainly due to the fact that they can lead to a peculiar pattern of non-Gaussian spectrum of the curvature perturbation~\cite{Alishahiha:2004eh,DBI:nG}.

In this paper we mainly focus on the slow-roll inflation. We show that an expansion of more than 60 $e$-foldings can be obtained in this model, by again realizing inflection point inflation. Namely, a D3-brane can slow-roll along the radial direction of the throat, spanning for a sufficiently long range before the angular directions become tachyonic. The amplitude and spectral tilt of the power spectrum of the curvature perturbation, produced during the epoch of slow-roll inflation, are compatible with the present data. The instability develops along the transverse angular directions when the inflaton reaches a region near the tip of the throat, and may make inflation ending \`{a} la hybrid inflation~\cite{Copeland:1994vg,hybrid}. All the process is fully under control from the supergravity point of view. We present the necessary tools to follow the dynamics explicitly, and estimate the masses of the fields involved along the inflationary trajectory. For our particular model of inflation, we find that angular directions become very light, and eventually tachyonic, only towards later stages of inflation. The angular directions then roll down the potential towards their true minima. The fact that the dynamics remain single field for most of the inflationary trajectory renders the predictions of our particular model similar to the ones of standard slow-roll inflation. On the other hand, the mass of the angular mode during inflation is never significantly larger than the Hubble scale. This can render the instability process particularly slow, with important observational consequences for the power spectrum as discussed in other contexts in Ref.~\cite{longhybrid} (see also Ref.~\cite{waterfall_pert} for recent analysis of the dynamics of cosmological perturbations due to the waterfall field in hybrid inflation).

To conclude this introduction, it is important to mention that hybrid inflation scenarios were considered in previous realizations of D-$\overline{\rm D}$ inflationary models, as discussed in some of the papers of Ref.~\cite{reprsSI}. In those contexts, inflation ends when an instability of the tachyon field connecting the D-$\overline{\rm D}$ system arises. This can lead to formation of cosmic strings at the end of inflation, which is a subject that has raised a large research activity starting with Ref.~\cite{braneinf:cosmicstring} (see for example Ref.~\cite{Polchinski:2004ia} for a review). An advantage of our model is that, as we mentioned, the development of instabilities along the angular directions can be described within a supergravity framework, without involving string theory tachyons. On the other hand, as we will discuss, after the instability process takes places the system does not immediately fall into a global minimum. Consequently, it is not possible to make definite statements about the production and stability of topological defects at the end of inflation in our setup without relying on more detailed analysis.

The paper is organized as follows. In Section~\ref{sec:conifold}, after introducing some basic facts of brane potentials in the warped deformed conifold, we present the explicit form of the potential in the Ouyang embedding. In Section~\ref{sec:angstaban}, we discuss the kinetic term of a mobile D3-brane and identify the properly normalized angular fields, and briefly address features of the angular masses near the tip of the throat. In Section~\ref{sec:numeric}, we show that, with an explicit example, that an expansion of 60 $e$-foldings of slow-roll inflation is possible in the Ouyang embedding. In Section~\ref{sec:multifield}, we give an  analysis of the multi-field potential of both radial and angular fields, and provide the necessary formulae for further study of the multi-field dynamics. In Section~\ref{sec:future}, we conclude concisely discussing interesting issues to be analysed in the future. We relegate calculational details to three appendices.

\section{Warped conifold as an inflationary playground}
\label{sec:conifold}

\subsection{Parametrizing the potential and multiple field inflation}
\label{sub-parametr}

In generic string compactifications, a smooth warped deformed conifold throat can easily develop in a bulk compact Calabi-Yau manifold near the conifold singularity. Such a deformation was done in the context of IIB string theory by the backreaction of localized imaginary self dual (ISD) three form fluxes~\cite{Giddings:2001yu}. To study the D3 trajectory in such a throat, an elegant systematic way to parametrize the potential experienced by a D3-brane was recently given in Refs.~\cite{Baumann:2010sx,Baumann:2008kq}, which can be summarized succintly as
\begin{equation}\label{VD3}
V_{\rm D3}=T_3 \Phi_{-} \, ,
\end{equation}
where $T_3=1/[(2\pi)^3(\alpha')^2]$ is the D3-brane tension, and $\Phi_-=e^A-\alpha$ is a combination of the throat warp factor and the five form field strength given in Ref.~\cite{Baumann:2008kq}. The mode $\Phi_{-}$ characterizes the perturbation away from the ISD background such as warped deformed conifold where $\Phi_-=G_-=0$. Here $G_{-}=(\star_6-i) G_3$ is the IASD component of the complex three form flux. At low orders in perturbative expansion, D3-branes only couple to $\Phi_-$ field, and do not couple to metric or dilaton fluctuations, thus giving the simple form (\ref{VD3}). The equation of motion of $\Phi_{-}$ can be derived explicitly from IIB supergravity action~\cite{Baumann:2010sx} as
\begin{equation}\label{EOMPhim}
\nabla^2\Phi_- = \frac{e^{8A+\phi}}{24}|G_{-}|^2 + {\mathcal R}_4 + e^{-4A}|\nabla \Phi_-|^2 + {\mathcal S}_{\rm local} \, ,
\end{equation}
where $e^{\phi}=g_s$ is the string coupling, $\calR_4$ is the four dimensional Ricci scalar and $S_{\rm local}$ denotes the localized sources. The Laplacian operator $\nabla^2$ is defined with respect to the unperturbed background metric. As the warped throat is attached to the bulk Calabi-Yau at some large radius $r_{\rm UV}$, $\Phi_{-}$ captures the bulk perturbations such as distant supersymmetry breaking fluxes or quantum effects. In the language of holography, at these large radius, ultraviolet perturbations are packaged into the so-called ``{\em non-normalizable modes}''. Moreover, $\Phi_-$ can also receive contributions from the ``{\em normalizable modes}'' of supergravity fields, which encode small radius, infrared perturbations. A particular example is the perturbation due to the $\overline{{\rm D3}}$ at the tip of warped deformed conifold which induces the D3-$\overline{{\rm D3}}$ Coulomb attraction~\cite{DeWolfe:2008zy}. In other words, $\Phi_-$ can in principle parametrize the potential for the entire trajectory of a D3-brane in the warped deformed conifold throat.

The parametric solution to (\ref{EOMPhim}) have been extensively investigated in Refs.~\cite{Baumann:2010sx,Baumann:2008kq}. The results can be expressed in a simple form as
\begin{equation}\label{paramVD3}
V_{\rm D3}(\phi)=V_{\rm D3\text{-}\overline{D3}}(\phi)+\sum_{i} c_i H_i(\Psi) \phi^{\Delta_i} \, ,
\end{equation}
where $c_i$ are constants and $\phi$ is the canonically normalized scalar field describing the radial motion of the D3-brane in the deformed conifold. The parameter $\Psi=\{\theta_1,\theta_2,\phi_1,\phi_2,\psi\}$ collectively denotes the five different angular coordinates of the conifold\footnote{See Appendix~\ref{app:conifold} for the explicit metric and angular coordinates of warped deformed conifold.}, and $H_i(\Psi)$ is expressible in terms of the angular harmonic functions. We have isolated the D3-$\overline{\rm D3}$ Coulomb interaction as it is a normalizable perturbation, while the polynomial series consists of the contributions from non-normalizable perturbations. It should also be noted that there can be additional $\phi$-independent constants added to $V_{\rm D3}(\phi)$ due to the coupling with the four dimensional Ricci scalar ${\mathcal R}_4$ which can also receive contributions from distant sources. In Refs.~\cite{Baumann:2010sx,Baumann:2008kq}, much of effort was devoted to enumerating the discrete spectrum of the scaling dimensions $\{\Delta_i\}$. The angular harmonic function $H_i(\Psi)$ is related to $\Delta_i$ and can also be computed in principle from the expansion of (\ref{EOMPhim})\footnote{To be more specific, the computations in Refs.~\cite{Baumann:2010sx,Baumann:2008kq} were done in the asymptotic, singular conifold limit. To extend the analysis into the deformed conifold region, we expect logarithmic corrections to the scaling dimensions $\{\Delta_i\}$, and modifications to the angular harmonic function $H_i(\Psi)$ as the $U(1)$ subgroup of the $SU(2)\times SU(2) \times U(1)$ singular conifold isometry group is broken down to discrete ${\mathbb Z}_2$.}. The values of the expansion coefficients $c_i$ however are model dependent, and can only be specified when quantities such as particular moduli stabilization effects, supersymmetry breaking fluxes and other microscopic quantities are known.

The parametrization (\ref{paramVD3}) makes it apparent that the radial and angular motions of a D3-brane are coupled in warped throat, and offer rich landscape for inflationary model building. On one hand, to obtain a single field inflationary model where $\phi$ plays the role of the canonical inflaton, it is necessary to restrict to a special trajectory in the parameter space where all the angular modes are either stabilized at their minima or decoupled from $\phi$: a good example was presented in Ref.~\cite{Baumann:2007ah}. We, on the other hand, should expect multi-field inflation being rather generic in the warped throat. If we are to realize this and construct an explicit potential, it is necessary to study how D3 angular modes $\Psi$ couple to the radial mode $\phi$. In particular, as we demonstrate in an example in Section~\ref{sec:numeric} after taking into account moduli stabilization, the angular masses can change sign for certain values of $\phi$, and become tachyonic. We have thus a situation where single field inflation is connected to a system with multiple field dynamics.

\subsection{An explicit inflaton potential: Ouyang embedding case}

We consider the motion of a D3-brane on a warped deformed conifold, whose geometrical properties are summarized in Appendix~\ref{app:conifold}. One of the basic quantities that characterize the geometry, and that plays an important role in what follows, is the deformation parameter $\epsilon$. We include a $\overline{\text{D3}}$-brane at the tip of the cone, and we take into account stabilization effects that are needed for providing masses to undesired light moduli. In this setup, we calculate the potential experienced by the moving D3-brane. In this and the following sections we mostly present our main results, while details can be found in the appendices to which we will refer in due course. 
 In order to calculate the D3-brane potential, we do not directly solve 
 (\ref{EOMPhim}). Instead, we follow a supergravity approach, that 
 allows us to apply well established techniques and results on the 
 properties of the warped deformed conifold with embedded D7-branes.

More specifically, we work in the framework of the KKLT moduli stabilization mechanism~\cite{Kachru:2003aw}, where in order to stabilize the K\"ahler moduli
\begin{equation}
\rho=\sigma+i\chi \, ,
\end{equation}
whose real part corresponds to the overall volume of the compact Calabi-Yau space, we consider the superpotential $W(z^\alpha,\rho)$ consisting of two contributions
\begin{equation}\label{eq:SuperW}
W(z^\alpha,\rho) =  W_0 + A(z^{\alpha})e^{-a\rho} \, .
\end{equation}
Here, $z^1,\cdots z^4$ parameterize the complex coordinates of the internal manifold. The first term $W_0=\int G_3\wedge \Omega_3$ is the perturbative Gukov-Vafa-Witten flux superpotential~\cite{Gukov:1999ya}, which, at least in principle, can stabilize the complex structure moduli and the dilaton-axion combination. Without loss of generality, we shall assume $W_0 \in \mathbb{R}^-$. It is well known that the tree level K\"ahler potential for $\rho$ exhibits no scale structure and leaves $\rho$ unfixed. A mechanism for stabilizing $\rho$ is therefore to include the non-perturbative gaugino condensate on a stack of space-filling D7-branes (or a Euclidean D3-brane) wrapping a holomorphic four cycle in the Calabi-Yau space, as appears in the second term of (\ref{eq:SuperW}). In the presence of a mobile D3-brane, the one-loop determinant $A(z^\alpha)$ picks up dependence on the D3-brane position moduli, which appear through the holomorphic D7 embedding function $f(z^\alpha) = 0$~\cite{Baumann:2006th}\footnote{The complex coordinates defining the deformed conifold are given in Appendix~\ref{app:conifold}.}. We can rewrite $A(z^\alpha)$ as
\begin{equation}\label{DefAz}
A(z^\alpha) = A_0 \left[ \frac{f(z^\alpha)}{f(0)} \right]^{1/n} \, .
\end{equation}
Here $A_0$ is a complex constant whose exact value depends on other stabilized complex structure moduli, and $n$ is the number of D7s (or $n=1$ for a Euclidean D3) giving the gaugino condensate (or instanton correction), which also enters in the definition of the exponent $a$ in (\ref{eq:SuperW}) as $a=2\pi/n$.
In other words, the non-perturbative gaugino condensate on D7-branes not only stabilizes K\"ahler modulus $\rho$, but also generates potential for the mobile D3-brane, echoing an earlier result of Ref.~\cite{Ganor:1996pe}. This also fits well with the general parametrization (\ref{paramVD3}), as it was also discussed in Ref.~\cite{Baumann:2010sx} that gaugino condensate on D7-branes can act as local sources for IASD flux, and make contribution to $\Phi_-$.

In this paper we focus on the so-called Ouyang embedding in warped deformed conifold~\cite{Ouyang:2003df,Chen:2008jj},
\begin{equation}\label{Defwembedding}
f(w_1)=\mu-w_{1} \, ,
\end{equation}
where $w_{1}=(z_1+i z_2)/\sqrt{2}$ and $\mu$ is a complex constant\footnote{The quantities $w_i$ ($i=1,\cdots4$) correspond to an alternative coordinate system for the internal manifold under investigation. For the relation between $\{w^{\alpha}\}$ and $\{z^{\alpha}\}$, see Appendix~\ref{app:conifold}.}. Note that $|\mu|$ heuristically measures the depth which D7-branes enter the conifold throat. The D7 embedding (\ref{Defwembedding}) breaks the $SU(2)\times SU(2)\times U(1)$ isometry of the singular conifold down to\footnote{The residual $U(1)\times U(1)$ isometry gets further broken to $U(1)$ in the deformed conifold. For the subtleties of supersymmetric D7 embeddings in warped deformed conifold, see Ref.~\cite{Chen:2008jj}.} $U(1)\times U(1)$, and can generate potential for the D3 angular modes associated with the broken isometries. The explicit potential for the $w$-embedding (\ref{Defwembedding}) can be obtained by substituting (\ref{eq:SuperW}), (\ref{DefAz}) and (\ref{Defwembedding}), along with the warped K\"ahler potential~\cite{warpedKahler} describing the kinetic terms for $\rho$ and D3-brane
\begin{equation}\label{eq:KahlerPotential}
\kappa^2{\cal K}( z^\alpha, \bar{z}^{\alpha},\rho,\bar{\rho}) =  -3 \log \left[ \rho+\bar{\rho}-\gamma k \left( z^\alpha,\bar{z}^\alpha \right) \right] \equiv -3\log U(\rho,\bar{\rho},z^\alpha,\bar{z}^\alpha)~,
\end{equation}
into the standard ${\mathcal N}=1$ supergravity expression of the $F$-term scalar potential. Here $\kappa^2 = \mpl^{-2} = 8\pi G$, $\gamma=\sigma_0 T_3/(3\mpl^2)$ with $\sigma_0$ being the stabilized value of $\sigma$ at the tip of the warped deformed conifold, and $k(z^\alpha,\bar{z}^\alpha)$ is the geometric K\"ahler potential for the deformed conifold given in (\ref{DefKahdefcon}). The detailed calculations are  similar to those for the Kuperstein embedding~\cite{Kuperstein:2004hy} as given in Refs.~\cite{Baumann:2007ah,Chen:2008ai}, and required inverse deformed conifold metric is given in (\ref{invmetric}), (\ref{Rij}) and (\ref{Lij}). After some calculations, we can obtain $V_F$ as
\begin{equation}\label{VF_before_angular_stab}
V_F = \mathcal{A}\mathcal{G}^{1/n} \left[ \mathcal{B} - 6ae^{a\sigma} \frac{|W_0|}{|A_0|}\mathcal{G}^{-1/(2n)} \right] + \frac{\mathcal{A}}{n} \mathcal{G}^{1/n-1} \left( \frac{k'}{k''}\coth\tau\mathcal{C} + \frac{\epsilon^2
\cosh{\tau}
}{a\gamma n\mu^2k''} \mathcal{D} \right) \, ,
\end{equation}
where
\begin{align}
\mathcal{A} \equiv & \frac{a\kappa^2e^{-2a\sigma}|A_0|^2}{3U^2} \, ,
\\
\mathcal{B} \equiv & a \left( U + \gamma \frac{{k'}^2}{k''} \right) + 6 \, ,
\\
\mathcal{C} \equiv & \frac{w_1+\overline{w_1}}{\mu} - 2\frac{|w_1|^2}{\mu^2} + \left( \frac{w_2+\overline{w_2}}{\mu} - \frac{w_1w_2+\overline{w_1}\overline{w_2}}{\mu^2} \right) \mathrm{sech}\tau \, ,
\\
\mathcal{D} \equiv & 1 - \frac{|w_2|^2}{\epsilon^2\cosh\tau} + \coth\tau \left( \frac{k''}{k'} - \coth\tau \right) \left[ \tanh^2\tau - \frac{w_1w_2+\overline{w_1}\overline{w_2}}{\epsilon^2\cosh^2\tau} - \frac{|w_1|^2+|w_2|^2}{\epsilon^2\cosh\tau} \right] \, ,
\\
\mathcal{G} \equiv & 1 - \frac{w_1+\overline{w_1}}{\mu} + \frac{|w_1|^2}{\mu^2} \, .
\end{align}
We can see that $V_F$ depends on the volume modulus $\sigma$, the radial coordinate $\tau$ and various combinations of the D3-brane coordinates $\left\{w_1+\overline{w_1}, |w_1|^2, w_2+\overline{w_2}, |w_2|^2, w_1w_2+\overline{w_1}\overline{w_2}\right\}$. The coordinate $\tau$ is associated with the radial coordinate $r$, frequently discussed when analysing the singular warped throat, by $r^3 = \epsilon^2 \cosh{\tau}$: see (\ref{Defr}). In the limit $A(w_1)\to A_0$, $V_F$ reduces to the scalar potential considered in Ref.~\cite{Kachru:2003sx}. We stress that similar computations have been done in the singular conifold limit of (\ref{Defwembedding})~\cite{Burgess:2006cb,Krause:2007jk}. Here we complete the computation for the full warped deformed conifold, as it will be crucial for studying the angular instability at small radius in a fully under control setup.

As the isometries of the conifold are partially broken by the D7 embedding (\ref{Defwembedding}), some of the angular directions can pick up effective masses through $V_F$, and their values are localized along the extremal trajectory which satisfies
\begin{equation}\label{extTraj}
\frac{\partial V_{F}}{\partial \Psi_i}=0 \, .
\end{equation}
In writing the above equation, we separate the radial coordinate $\tau$ from the five angular coordinates of the warped conifold, collectively denoted with $\Psi_i$. For the embedding (\ref{Defwembedding}), two such trajectories exist and are given by
\begin{align}
\label{wextTraj}
& w_1^{(0)}=\pm\frac{\epsilon}{\sqrt{2}}e^{\tau/2} \, , \qquad w_2^{(0)}=\mp\frac{\epsilon}{\sqrt{2}}e^{-\tau/2} \, , \qquad \text{(non-delta-flat)}
\\
\label{wextTraj2}
& w_1^{(0)}=\pm\frac{\epsilon}{\sqrt{2}}e^{-\tau/2} \, , \qquad w_2^{(0)}=\mp\frac{\epsilon}{\sqrt{2}}e^{\tau/2} \, . \qquad \text{(delta-flat)}
\end{align}
The detailed steps are given in Appendix~\ref{app:angular}. Notice that these extremal trajectories do not fix the overall sign of $w_1$ and $w_2$, but only the relative sign between the two quantities. The choice of the overall sign has important physical consequences that we will discuss in the following sections. The resultant $F$-term scalar potential along these extremal trajectories is given by
\begin{equation}\label{VFtausigma}
V_F(\tau,\sigma) = \calA(\tau,\sigma) \calG_0^{1/n}(\tau) \left[ \calB(\tau,\sigma) -
6ae^{a\sigma}\frac{|W_0|}{|A_0|}\calG_0^{-1/(2n)}(\tau) + \calF(\tau) \right] \, ,
\end{equation}
where
\begin{align}
\calA(\tau,\sigma) = & \frac{a\kappa^2e^{-2a\sigma}|A_0|^2}{3U^2(\tau,\sigma)} \, ,
\\
U(\tau,\sigma) = & 2\sigma - \frac{3^{1/3}}{2^{2/3}an}\beta \int_0^\tau d\tau' \left[ \sinh(2\tau')-2\tau' \right]^{1/3} \, ,
\\
\calB(\tau,\sigma) = & a \left\{ U(\tau,\sigma) + \frac{3^{4/3}}{2^{8/3}}\frac{\beta}{an} \frac{[\sinh(2\tau)-2\tau]^{4/3}}{\sinh^2\tau} \right\} + 6 \, ,
\\
\calF(\tau) = & \frac{1}{n\calG_0(\tau)} \left[ \frac{3}{4} \frac{\sinh(2\tau)-2\tau}{\sinh^2\tau}\coth\tau\calC_0(\tau) + \frac{3^{2/3}}{2^{1/3}} \frac{\alpha^2}{\beta} \frac{\cosh\tau}{\sinh^2\tau} \left[ \sinh(2\tau)-2\tau \right]^{2/3}\calD_0(\tau) \right] \, ,
\end{align}
and the functions which explicitly depend on the choice of the extremal trajectories are
\begin{align}
\calC_0(\tau) = &
 \begin{cases}
  \pm2\alpha e^{\tau/2}\tanh\tau \sqrt{\calG_0(\tau)} \, , & \text{(non-delta-flat)}
  \\
  \mp2\alpha e^{-\tau/2}\tanh\tau \sqrt{\calG_0(\tau)} \, , & \text{(delta-flat)}
 \end{cases}
\\
\calD_0(\tau) = &
 \begin{cases}
  \cfrac{e^{\tau}}{2\cosh\tau} \, , & \text{(non-delta-flat)}
  \\
  \cfrac{e^{-\tau}}{2\cosh\tau} \, , & \text{(delta-flat)}
 \end{cases}
\\
\calG_0(\tau) = &
 \begin{cases}
  \left(1 \mp \alpha e^{\tau/2} \right)^2 \, , & \text{(non-delta-flat)}
  \\
  \left(1 \mp \alpha e^{-\tau/2} \right)^2 \, . & \text{(delta-flat)}
 \end{cases}
\end{align}
Here, we have defined two dimensionless parameters $\alpha$ and $\beta$ by
\begin{align}
\label{def:alpha}
\alpha \equiv & \frac{\epsilon}{\sqrt{2}\mu} \, ,
\\
\label{def:beta}
\beta \equiv & 2\pi\gamma k''|_{\tau=0} = \frac{2^{1/3}}{3^{1/3}}an\gamma\epsilon^{4/3} \, .
\end{align}
They have the following geometrical meaning. $\alpha$ measures the depth which the D7-branes extend into the deformed conifold, while $\beta$ is inversely proportional to the four dimensional Planck mass $\mpl$, and hence to the ultraviolet cutoff $r_\text{UV}$ of the warped throat, i.e. $\beta \sim \epsilon^{4/3}/r_\text{UV}^2$. We therefore deduce that $r_\text{UV} \gtrsim \mu^{2/3}$, and equivalently $\alpha \gtrsim \beta$. In fact, when we take into account the contribution to $\mpl$ due to the compact Calabi-Yau to which the throat is connected, we generally expect $\alpha \gg \beta$. Also, the sign choice depends on the sign chosen in the extremal trajectories. Choosing the upper sign implies to choose the upper sign in the expressions above, and vice versa. In obtaining (\ref{VFtausigma}), we have also stabilized the axion $\chi$ in the K\"ahler modulus $\rho$. The scalar potential now appears as a function of only two variables $\tau$ and $\sigma$, which we also necessarily need to stabilize to prevent decompactifcation.

Before we conclude this section, let us  make some remarks. First, as noted in Ref.~\cite{Baumann:2007ah}, the non-delta-flat trajectory represents an unstable minimum. While at large radius, the angular mass matrix normally has positive eigenvalues, at small radius the trajectory becomes unstable and the angular modes become light and tachyonic. In the next sections we shall investigate in detail such instability and the multi-field potential. Second, we should note that in addition to the $F$-term scalar potential, the mobile D3-brane also experiences a Coulomb attraction coming from the $\overline{\text{D3}}$ at the tip of warped deformed conifold, whose form is given by
\begin{equation}\label{VD3D3bar}
V_{\textrm{D3}\text{-}\overline{\rm D3}}(\tau,\sigma) = \frac{D_0}{U^2(\tau,\sigma)} \left[ 1-\frac{3D_0}{16\pi^2 T_3^2(\Delta y)^4} \right] \, ,
\end{equation}
where $D_0=2T_3a_0^4 $ and $\Delta y$ parameterizes the D3-$\overline{\text{D3}}$ separation. This provides an additional contribution to the inflationary potential. Notice that, as shown in 
 Ref.~\cite{Kachru:2003sx}, if the separation between brane and antibrane is
 sufficiently large, the Coulomb potential does not depend on the angular coordinates.
 Third, let us also briefly comment on the possible relevance of worldvolume fluxes on 
 the D7-branes that can in principle induce a force on the moving D3-brane. As shown in
 Ref.~\cite{Baumann:2007ah}, this force is generically suppressed with respect
 to the other contributions to the D3 potential. Consequently, worldvolume flux effects
 are not expected to qualitatively change the analysis of the inflationary dynamics that
 we are going to discuss.

 To conclude, let us once again discuss the difference between the supergravity approach
 adopted in this section and in the rest of the paper, and the more direct approach
 introduced in the previous section based on the dynamics of the field $\Phi_{-}$. 
 The two approaches should provide the same results, as long as one neglects
 the effects of the compact Calabi-Yau manifold to which the throat is attached at large 
 radius. The method of Section~\ref{sub-parametr} is in principle suitable also for
 taking into account the effects of the bulk Calabi-Yau that are instead hard to describe
 within the supergravity approach of this section. On the other hand, as we will discuss 
 in due course, the phenomena that we analyse in this paper are not particularly sensitive 
 to these bulk effects, and our results are expected to remain correct also when we include
 these contributions.

\section{Angular stability analysis}
\label{sec:angstaban}

Naively, after we further stabilize  the volume modulus $\sigma$ in (\ref{VFtausigma}) at a given value of $\tau$, we obtain a single field potential in $\tau$. However for this statement to hold for the entire deformed throat, we need to ensure that the broken angular isometries are much heavier than the radial mode and effectively stabilized at their minima along the entire extremal trajectories. In this section, we investigate this issue for all broken isometries by examining their effective masses which depend on the radial direction. The first step is to identify the canonically normalized fields associated with the angular coordinates. As we discuss in Appendix~\ref{app:angular}, when considering small displacements of the angular directions from the extremal trajectories, it is convenient to assemble the angular degrees of freedom into three variables that we call $P$, $Q$ and $R$.
 We concentrate on fields representing small displacements from the extremal
 trajectory since, as we will discuss in more detail in what comes next, the 
 angular masses during slow-roll inflation are large and positive. So angular 
 modes remain localized nearby the extremal trajectory. Only at later stages 
 of slow-roll inflation, as we will see, angular modes become light and large
 displacements from the extremal trajectory can occur. We will discuss the 
 consequences of this fact in Section~\ref{sec:multifield}.

The kinetic term of a mobile D3-brane can be deduced from the K\"ahler potential (\ref{eq:KahlerPotential}) as
\begin{equation}\label{D3KE}
{\mathcal L}_{\rm kin} = -\frac{T_3}{2}\frac{\sigma_0}{\sigma_\star(\tau)} \, d\hat{s}_6^2 \, .
\end{equation}
Here, $d\hat{s}_6^2$ denotes the pull-back of the warped deformed metric (\ref{eq:DeformMetric}) and $\sigma_\star(\tau)$ is the solution to the stabilization condition
\begin{equation}
\left. \frac{\partial\left(V_F+V_{\rm D3\text{-}\overline{D3}}\right)}{\partial\sigma} \right|_{\sigma=\sigma_\star} = 0 \, .
\end{equation}
Note that using $\sigma_\star$, we can write $\sigma_0=\sigma_\star(\tau=0)$. We can obtain the analytic expression of $\sigma_\star(\tau)$ for small $\tau$ region as in the case of the Kuperstein embedding: see Appendix~\ref{secstabsigma}. But for general $\tau$ we have to solve the stabilization condition numerically. Considering small angular displacements around the extremal trajectories, the pull-back of the metric is given by (\ref{ExpAngKE}) as
\begin{equation}\label{ExpAngKE1}
\left. d\hat{s}^2_6 \right|_{0} =
\frac{\epsilon^{4/3}}{2}K(\tau) \left\{ \frac{d\tau^2+4dP^2}{3K^3(\tau)} + 4 \left[
\cosh^2\left( \frac{\tau}{2} \right) dR^2 + \sinh^2\left( \frac{\tau}{2} \right) dQ^2 \right] \right\} \, .
\end{equation}
From this form, we can canonically normalize the radial direction by defining a scalar field $\phi(\tau)$ as
\begin{align}\label{canfi}
\phi(\tau) \equiv & \sqrt{\frac{T_3}{6}}\epsilon^{2/3} \int_0^{\tau}\,\frac{d\tau'}{K(\tau')}\,\sqrt{
\frac{\sigma_0}{\sigma_\star(\tau')}}
\nonumber\\
= & \frac{3^{1/6}}{2^{2/3}} \sqrt{\frac{\beta}{an\sigma_0}} \mpl \int_0^{\tau}\,\frac{d\tau'}{K(\tau')} \, \sqrt{\frac{\sigma_0}{\sigma_\star(\tau')}} \, ,
\end{align}
where the function $K(\tau)$ is given by (\ref{DefKtau}). Also we rescale angular fluctuations including constant factors as
\begin{align}
\widehat{P} \equiv & \sqrt{\frac{2}{3}T_3}\,\epsilon^{2/3}\,P
= 2^{1/3}3^{1/6} \sqrt{\frac{\beta}{an\sigma_0}} \mpl \,P
\, ,
\\
\widehat{R} \equiv & \sqrt{2\,T_3}\,\epsilon^{2/3}\,R
= 2^{1/3}3^{2/3} \sqrt{\frac{\beta}{an\sigma_0}} \mpl \,R
\, ,
\\
\widehat{Q} \equiv & \sqrt{2\,T_3}\,\epsilon^{2/3}\,Q
= 2^{1/3}3^{2/3} \sqrt{\frac{\beta}{an\sigma_0}} \mpl \,Q
\, ,
\end{align}
so that they pick up mass dimension 1. We can then write the kinetic term Lagrangian as
\begin{equation}
{\mathcal L}_{\rm kin} = -\frac{1}{2} \left(\partial\phi \right)^2 - \frac{K(\tau)}{2}\frac{\sigma_0}{\sigma_\star(\tau)} \left[ \frac{\left(\partial\widehat{P}\right)^2}{K^3(\tau)} + \cosh^2\left( \frac{\tau}{2} \right) \left(\partial\widehat{R}\right)^2 + \sinh^2\left( \frac{\tau}{2} \right) \left(\partial\widehat{Q}\right)^2 \right] \, ,
\end{equation}
where $\tau$ should be regarded as a function of $\phi$ by inverting relation (\ref{canfi}), i.e. $\tau=\tau(\phi)$. Kinetic terms with field-dependent coefficients, as the previous ones, have been used in the past in the context of multiple field inflation. They arise when discussing models characterized by a
non-trivial metric in field space: see Ref.~\cite{generalmulti} for some literature. We will use this form  of kinetic terms in our numerical analysis in Section~\ref{sec:numeric}.

Let us now discuss the signs of the mass eigenvalues of the angular fields $P$, $Q$ and $R$ in the small $\tau$ limit. A similar analysis was performed in Refs.~\cite{Baumann:2007ah,Burgess:2006cb} in the singular conifold limit, and the results showed that some of the angular directions become generically tachyonic at small values of the radial direction. Here we complete the analysis to the case of warped deformed conifold, and analyse in detail the range of parameters leading to instabilities.

In the limit $\tau \to 0$, the two different extremal trajectories, (\ref{wextTraj}) and (\ref{wextTraj2}) coincide. After some calculations (see Appendix~\ref{app:angular} for details), we can find that the mass eigenvalues for the angular modes $P$, $Q$ and $R$, that we denote respectively with $X_P$, $X_Q$ and $X_R$, are given by
\begin{align}
X_P = & \mp2\alpha \frac{\calA\calG_0^{1/n-1}}{n} \left[ 2s - 3 + \frac{\alpha^2}{\beta} (1\mp\alpha)^{-2} \right] \, ,
\\
X_Q = & 0 \, ,
\\
X_R = & 2(1\mp\alpha)X_P + 4\alpha^2 \frac{\calA\calG_0^{1/n-1}}{n}\left( \frac{6}{5\beta} - 1 \right) \, .
\end{align}
At this point, we can understand whether the eigenvalues are negative or positive, when approaching the tip of the cone ($\tau \to 0 $). Instead of doing the full systematic analysis, here we concentrate on one representative case that allows us to qualitatively understand under which conditions an instability can arise. By choosing the lower sign in $X_P$, with uplifting ratio $s$ (\ref{Defs}) being large enough, it is easy to make sure that both $X_P$ and $X_R$ are positive at the tip of the warped throat\footnote{Notice
that this does not imply that $X_P$ and $X_R$ are always positive at $\tau >0$.}. We focus here on this case, that will be the one analysed in more detail in the following sections. On the other hand, since $X_Q\to 0$ as $\tau \to 0$, suppose it approaches zero from below: since  it is positive at large $\tau$, by continuity, this means that it changes sign in an intermediate region, and develops an instability. A sufficient condition to ensure that the eigenvalue approaches zero from below is to demand that it has negative first derivative along $\tau$, in the limit $\tau\to0$. A simple calculation provides
\begin{equation}
\left. \frac{\partial X_Q}{\partial\tau} \right|_{\tau=0} = \frac{2\alpha\,{\cal A}\, {\cal G}^{1/n-1}}{n} \,\left[ \left(2 s -3 \right) \left(1+\alpha \right) -\alpha-\frac{4 \alpha}{5 \beta (1+\alpha)}\left(1-\frac{\alpha}{5}\right) \right] \, .
\end{equation}
When the ratio $\alpha/\beta$ is sufficiently large, this is negative, but $\alpha/\beta$ is generally much larger than one: see the discussion after (\ref{def:beta}). We see explicitly that $X_Q$ becomes negative for sufficiently small $\tau$ while it vanishes at $\tau=0$. These analytic considerations will be confirmed by the numerical analysis of the following section.

\section{Slow-roll inflation in the Ouyang embedding}
\label{sec:numeric}

In the previous section, we found that there appears an instability at least in one or more angular masses along the extremal trajectories for natural choices of parameters. On the other hand, by suitably tuning the parameters, it might be  possible to realize slow-roll inflation that lasts for more than 60 $e$-folds well before the instability develops. In this section, we show that this is indeed the case. We specifically focus on the lower branch solution of the non-delta-flat direction (\ref{wextTraj}), i.e. $w_1=-\epsilon e^{\tau/2}/\sqrt{2}$ and $w_2 = \epsilon e^{-\tau/2}/\sqrt{2}$. The discussion on the application of the delta-flat trajectory to the multi-field inflation is postponed in the next section. The choice of the parameters is summarized in Table~\ref{Table:choice}.
\begin{table}[h]
 \begin{center}
\begin{tabular}{c*{5}{c}}
 \hline\hline
 $\alpha$ & $\beta$ & $n$ & $s$    & $|W_0|$                   & $A_0$
 \\
 0.00444  & 0.00157 & 8   & 1.1254 & $4.64\times10^{-6}\mpl^3$ & $0.008\mpl^3$
 \\
 \hline\hline
\end{tabular}
 \end{center}
 \caption{The parameters used in the numerical computations.}
 \label{Table:choice}
\end{table}

\begin{figure}[ht]
 \begin{center}
  \includegraphics[width = 15cm]{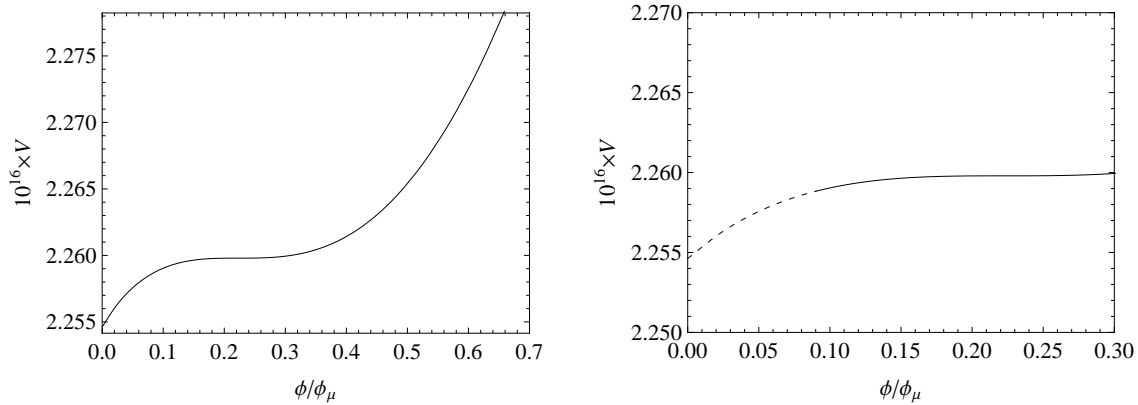}\vspace{-1.5em}
 \end{center}
 \caption{The full potential versus normalized canonical field $\phi/\phi_\mu$, which runs from 0 to 1. We set $\mpl = 1$. In the left panel, we simply show the entire potential, while in the right panel we concentrate on the inflecting region, where after the instability of $Q$ develops we show the potential with a dotted line.}
 \label{Fig:potential}
\end{figure}

\begin{figure}[h]
 \begin{center}
  \includegraphics[width = 15cm]{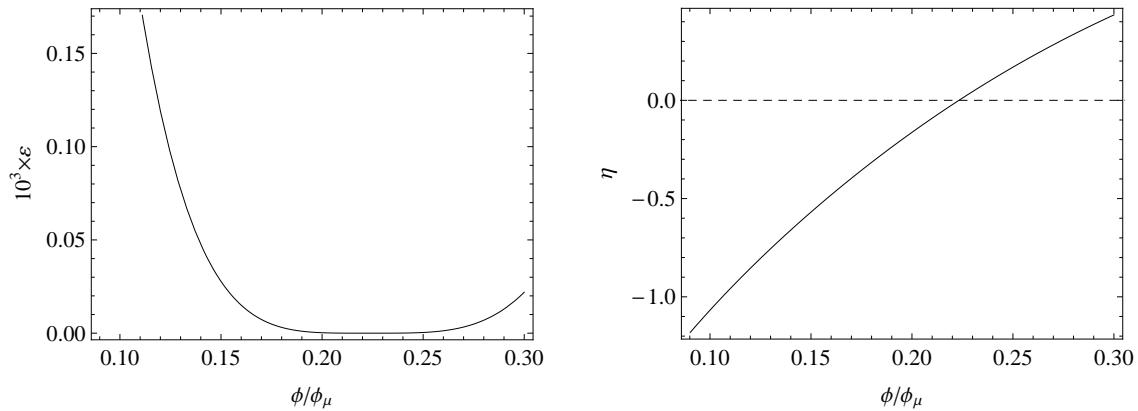}\vspace{-1.5em}
 \end{center}
 \caption{The ``potential'' slow-roll parameters (left) $\varepsilon$ and (right) $\eta$. Note that while $|\eta| \sim 1$ when the instability of $Q$ develops, $\varepsilon$ remains smaller than 1: at the moment of instability, we have $\varepsilon \approx 3.58931 \times 10^{-4}$.}
 \label{Fig:SR}
\end{figure}

The single field potential along the radial direction $\tau$ is shown in Fig.~\ref{Fig:potential} in terms of the canonically normalized field $\phi(\tau)$. Here we plot the potential in terms of $\phi(\tau)/\phi_{\mu}$, where
\begin{equation}
\phi_{\mu}^2 \equiv \frac{3}{an\sigma_0} \left( \frac{3}{2\alpha} \right)^{4/3}\beta \mpl^2 \, .
\end{equation}
$\phi_\mu$ corresponds to the maximal value for $\phi$ in the warped throat. We see that there is a point where the second derivative of the potential with respect to $\phi$ vanishes. Slow-roll inflation can be realized around this {\em inflection point}. In order to understand whether inflation can be realized or not, we define the slow-roll parameters\footnote{Notice that we write the first slow-roll parameter as $\varepsilon$ to distinguish it from the deformation parameter $\epsilon$ used elsewhere.}
\begin{align}
\varepsilon \equiv & -\frac{\dot{H}}{H^2} \approx \frac{\mpl^2}{2} \left( \frac{V'}{V} \right)^2 \, ,
\\
\eta \equiv & -\mpl^2 \frac{V''}{V} \, .
\end{align}
These slow-roll parameters are shown in Fig.~\ref{Fig:SR}. As expected, $\varepsilon$ remains small but $|\eta|$ quickly becomes large once the field moves away from the inflection point.

\begin{figure}[h]
 \begin{center}
  \includegraphics[width = 15cm]{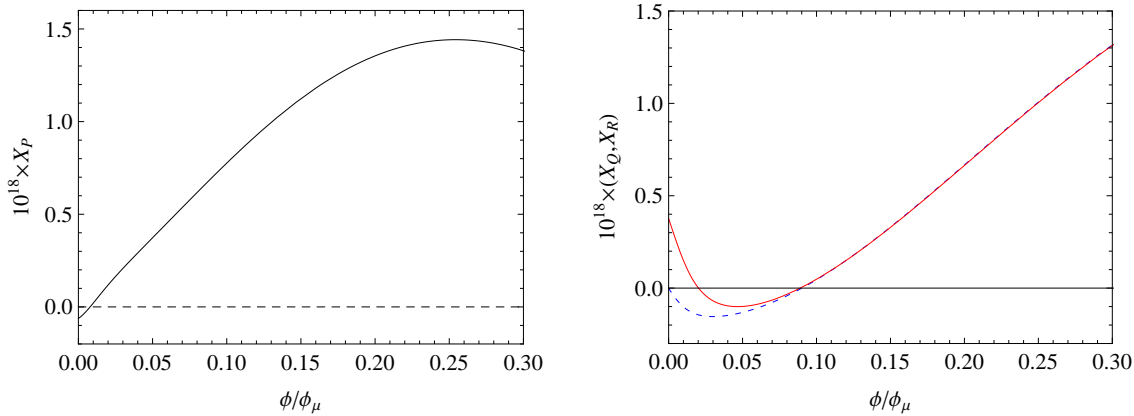}\vspace{-1.5em}
 \end{center}
 \caption{The angular masses (left panel) $X_P$, (right panel) (dotted line) $X_Q$ and (solid line) $X_R$. The difference between $X_Q$ and $X_R$ is only noticeable near the tip region.}
 \label{Fig:masses}
\end{figure}

Next we consider the masses of the angular directions. In Fig.~\ref{Fig:masses}, we show $X_P$, $X_Q$ and $X_R$. For our choice of parameters, we find that all the angular directions become unstable at small radius. However we should note that once instability develops, we cannot trust anymore the analysis of the mass matrix around the extremal trajectory in Appendix~\ref{app:angular}, as the angular displacements need not to be small since angular fields start to roll towards their true minimum. Thus we only study slow-roll inflation before one of the angular directions becomes unstable: in this case the first among the angular directions whose mass squared vanishes is $Q$. From this point on, we need to consider multi-field dynamics, which is the subject of the next section.

\begin{figure}[h]
 \begin{center}
  \includegraphics[width = 8cm]{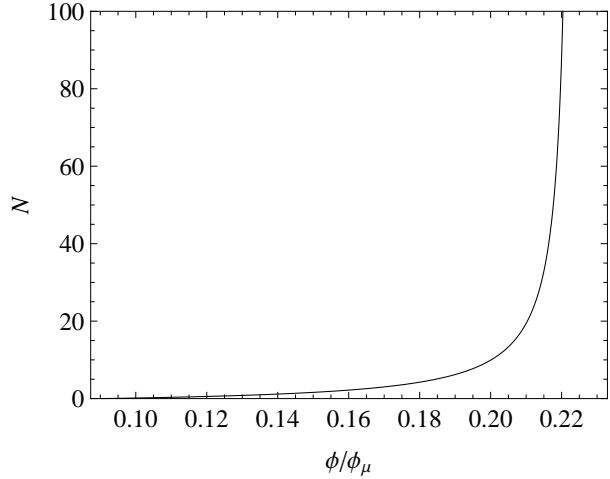}\vspace{-1.5em}
 \end{center}
 \caption{The number of $e$-folds $N$ versus $\phi/\phi_\mu$ {\em counted from the instability point}. As we approach the inflection point $N$ diverges, which indicates that we have indefinitely large $N$ near the inflection point.}
 \label{Fig:efold}
\end{figure}

We can estimate the number of $e$-folding number from this instability point as
\begin{equation}
N = \sqrt{\frac{3\beta}{2an\sigma_0}} \left( \frac{3}{2\alpha} \right)^{2/3} \int_{\phi_\mathrm{inst}}^\phi \frac{d\phi}{\sqrt{\varepsilon}} \, ,
\end{equation}
which is shown in Fig.~\ref{Fig:efold}. As we can see, the number of $e$-folds blows up near the inflection point. That is, we expect an indefinitely large number of $e$-folds if the classical initial condition is such that the field starts from very near the inflection point. Of course, in reality quantum fluctuations will push the field away from the inflection point. In any case, we learn that it is not difficult to realize slow-roll inflation that lasts for more than 60 $e$-folds before instability develops along one of the angular directions. An interesting, separated issue is the problem of initial conditions that can lead to inflation: we will not discuss it in this context, but see for example Ref.~\cite{overshoot}.

\begin{figure}[h]
 \begin{center}
  \includegraphics[width = 15cm]{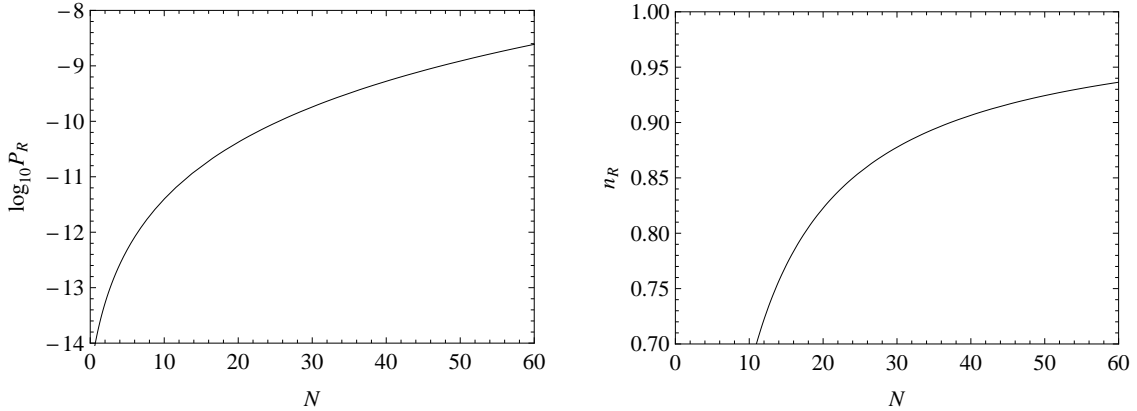}\vspace{-1.5em}
 \end{center}
 \caption{(Left panel) $\log_{10}\calP_\calR$ and (right panel) $n_\calR$ versus $N$. At $N = 60$, the values of $\calP_\calR$ and $n_\calR$ are close to the observed values.}
 \label{Fig:Pandn}
\end{figure}

Also, we calculate the power spectrum of the curvature perturbation ${\cal P}_{\calR}$ and the corresponding spectral index $n_\calR$, under the slow-roll approximation given by
\begin{align}
\calP_\calR = & \frac{V}{24 \pi^2 \varepsilon \mpl^4} \, ,
\\
n_\calR = & 1- 6 \varepsilon + 2 \eta \, .
\end{align}
In Fig.~\ref{Fig:Pandn}, we show $\log_{10}\calP_\calR$ and $n_\calR$ versus the number of $e$-folds. For our choice of parameters, we obtain $\calP_\calR = 2.44096 \times 10^{-9}$ and $n_\calR = 0.936381$ at $N=60$, which are well within the $2\sigma$ range of the current observations~\cite{Komatsu:2010fb}.

The observational predictions from slow-roll inflation is very similar to those in the Kuperstein embedding~\cite{Chen:2008ai}. This suggests that the inflection point inflation is quite generally realized regardless of the choice of the embeddings. In fact, as discussed in Section~\ref{sec:conifold}, it was suggested that the D3-brane potential can be expressed as
\begin{equation}
V= \sum_i c_i \phi^{\triangle_i} H_i (\Psi) \, ,
\end{equation}
where $\Psi$ is the angular coordinates, $c_i$ are constants and $\Delta_i$ are given by
\begin{equation}
\Delta_i = 1, \; \frac{3}{2}, \; 2, \; \frac{5}{2},\; 3,\; \cdots \, .
\end{equation}
In our setup, we find that the potential along the extremal trajectory is well fitted by the polynomial form
\begin{equation}
\frac{V}{\mpl^4} = V_0 \left[ 1 + 0.090 \left(\frac{\phi}{\mpl} \right) + 0.108 \left( \frac{\phi}{\mpl} \right)^{3/2} -2.637 \left( \frac{\phi}{\mpl} \right)^2 + 6.316 \left( \frac{\phi}{\mpl} \right)^{5/2} - 3.678 \left( \frac{\phi}{\mpl} \right)^3 \right] \, ,
\end{equation}
where $V_0 = 2.250 \times 10^{-16}$. This constant term arises from the uplifting contribution that is necessary to obtain a quasi de Sitter solution. This parametrization emphasizes the fact that inflation is obtained around an inflection point region, that results from a delicate cancelation among the various terms in the previous expansion. Notice that although the previous expression provides a potential for the canonically normalized inflaton along the radial direction $\phi$, the dependence on the angular coordinates is contained in the coefficients $H_i(\Psi)$. Their form can be deduced from the explicit scalar potential derived earlier, which also gives mass eigenvalues $X_P$, $X_Q$ and $X_R$ along the extremal trajectories. The knowledge of these quantities is necessary to understand whether or not the angular masses are heavier than the Hubble parameter during inflation: if not, angular directions are not stabilized on their extremal value and their dynamics can have interesting observational consequences.

\begin{figure}[h]
 \begin{center}
  \includegraphics[width = 15cm]{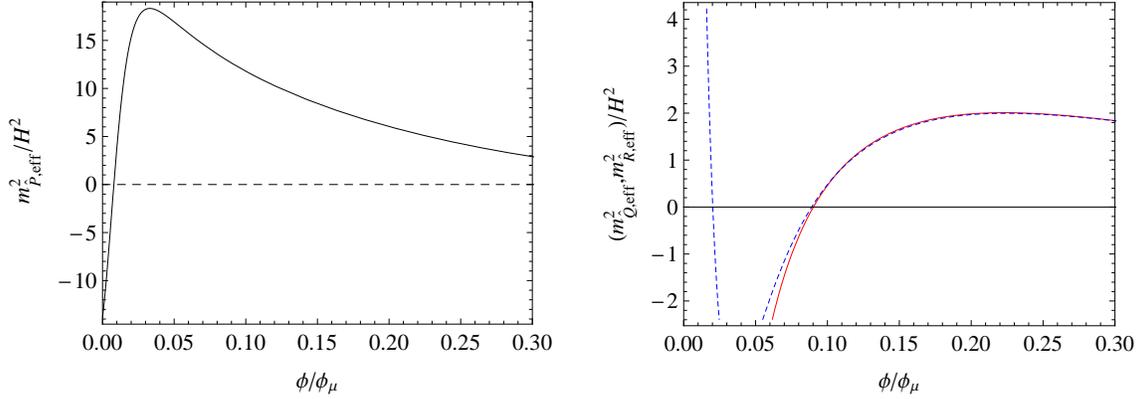}\vspace{-1.5em}
 \end{center}
 \caption{The plots of (left panel) $m_{\widehat{P},\text{eff}}^2/H^2$ and (right panel) (solid line) $m_{\widehat{Q},\text{eff}}^2/H^2$ and (dotted line) $m_{\widehat{R},\text{eff}}^2/H^2$ versus $\phi/\phi_\mu$. Note that we assume $H$ is completely dominated by the potential, which should be a good enough approximation near the flat region of the potential.
}
 \label{Fig:mass_ratios}
\end{figure}

To estimate the effective masses for the angular directions, as can be read from (\ref{ExpAngKE1}), we notice the kinetic terms induce a mixing between the radial field $\phi$ and the angular fields $\widehat{P}$, $\widehat{Q}$ and $\widehat{R}$. Precisely speaking, we need to properly take into account this non-trivial field space metric to analyze the masses of the angular fields and we will explain the procedure in more detail in the next section. However, during inflation, the motion along the radial direction is suppressed by the slow-roll parameter. Thus at the leading order in the slow-roll approximation, we can assume that the radial field is constant, as well as the coefficients of the kinetic terms. We are then interested in the {\em effective} masses, that take into account the non-trivial $\tau$-dependent coefficients in the kinetic term, as
\begin{align}
m_{\widehat{P},\text{eff}}^2 = & \frac{X_P}{\mpl^2}\,\frac{an}{2^{2/3}3^{1/3}\beta}\,\sigma_\star(\tau)\, K^2(\tau) \, ,
\\
m_{\widehat{Q},\text{eff}}^2 = & \frac{X_Q}{\mpl^2}\,\frac{an}{2^{2/3}3^{4/3}\beta} \, \frac{\sigma_\star(\tau)}{K(\tau) \sinh^2\left(\tau/2\right)} \, ,
\\
m_{\widehat{R},\text{eff}}^2 = & \frac{X_R}{\mpl^2}\,\frac{an}{2^{2/3}3^{4/3}\beta} \, \frac{\sigma_\star(\tau)}{K(\tau) \cosh^2\left(\tau/2\right)} \, .
\end{align}
For a given $\phi$, these masses characterize the behaviour of the isocurvature perturbations along the angular directions. We found at $N=60$, their ratios to the Hubble parameter are
\begin{align}
\frac{m_{\widehat{P},\text{eff}}^2}{H^2} = & 5.31529 \, ,
\\
\frac{m_{\widehat{Q},\text{eff}}^2}{H^2} = & 2.00828  \, ,
\\
\frac{m_{\widehat{R},\text{eff}}^2}{H^2} = & 1.99503 \, ,
\end{align}
and the behaviours of these ratios are shown in Fig.~\ref{Fig:mass_ratios}. We learn that the effective masses of the angular fields are comparable to the Hubble parameter during inflation: they are stabilized on their extremal values until the instability region is reached, at which the fields become tachyonic.

This conclusion is not unexpected. In supergravity, the typical mass of the moduli fields during inflation is of $\mathcal{O}(H)$. This is the reason of the $\eta$ problem that we mentioned in Section~\ref{sec:intro}: only by carefully tuning the parameters, one can obtain a flat potential for the inflaton field in some region of parameter space. In our framework, we can also understand more explicitly why the masses of the angular modes turn out to acquire these values during inflation, using the large $\tau$ expressions for $X_{P,Q,R}$. These quantities, in this limit, are found to be
\begin{align}
\label{large-tau-P}
X_P \to & \pm\frac{\mathcal{A}}{n} \left( 1 \mp \alpha e^{\tau/2} \right)^{-2(1/n-1)} \alpha e^{\tau/2}
\left\{ 2a\sigma + \frac{9}{2} - 3ae^{a\sigma} \frac{|W_0|}{|A_0|} \left( 1 \mp \alpha e^{\tau/2} \right)^{1/n} \right.
\nonumber\\
& \left. \hspace{2cm} - \left( 1-\frac{1}{n} \right) \left( 1 \mp \alpha e^{\tau/2} \right)^{-2} \left[ \pm 2\alpha e^{\tau/2} \left( 1 \mp \alpha e^{\tau/2} \right) + 3^{2/3}\frac{\alpha^2}{\beta}e^{\tau/3} \right] \right\} \, ,
\\
\label{large-tau-QR}
X_Q, X_R \to & \left( 1 \mp \alpha e^{\tau/2} \right)X_P + \frac{2\mathcal{A}}{3^{1/3}n} \frac{\alpha^2}{\beta} e^{\tau/3} \left( 1 \mp \alpha e^{\tau/2} \right)^{-2(1-1/n)} \left( -1 + \frac{3^{4/3}}{2} \beta e^{2\tau/3} \right) \, .
\end{align}
With our choice of parameters, the stabilisation condition for $\sigma$ can be approximated by (\ref{stabilisation}), which gives $\sigma_{\star} \sim \sigma_0$ even for large $\tau$. Then substituting (\ref{stabilisation}) into the above expressions, we find that the term proportional to $a \sigma$ cancels out where $a \sigma_{\star} \gg 1$ in order for the supergravity approximation to be valid. Comparing these expressions with the potential (\ref{VFtausigma}), we can estimate $X_{P,Q,R}$ as
\begin{equation}
X_{P,Q,R} =  \frac{{\cal N}_{P,Q,R}}{a n \sigma_{\star}}\,\, V,
\end{equation}
where ${\cal N}_{P,Q,R}$ are functions of $\alpha e^{\tau/2}$, $\alpha^2  e^{\tau/3}/\beta$ and $s$.
In the example we studied here, $\alpha e^{\tau/2} \ll 1, \alpha^2 e^{\tau/3}/\beta \ll 1$ and
${\cal N}_{P,Q,R} = \mathcal{O}(\alpha e^{\tau/2})$ at $N=60$. Then the effective masses can be estimated as
\begin{equation}\label{estmass}
\frac{m_{\widehat{P},\widehat{Q}, \widehat{R}, \text{eff}}^2}{H^2}
=\mathcal{O} \left(\frac{\alpha}{\beta} e^{-\tau/6} \right) \, .
\end{equation}
This is generally larger than one as $\alpha \gg \beta$, but they are not significantly large  with our choices of parameters: the mass eigenvalues during inflation turn out to be of the order of the Hubble scale. This can have important consequences for the field dynamics right at the end of inflation. Indeed,  in different contexts, mainly motivated by hybrid inflation, it has been shown that when the waterfall  fields have not too large masses the instability process might take longer than one Hubble time to complete~\cite{longhybrid}. This implies that slow-roll motion along the radial direction can continue, while the angular modes roll, not necessarily too rapidly, towards their true minima. It would be interesting to investigate these issues in the present context, although the analysis has to take into account the specific form of both the potential and the non-canonical kinetic terms.

 Let us finish this section discussing whether the results of this section
 can be modified after taking into account the effects of the bulk Calabi-Yau 
 attached to the throat at large radius. As shown in Ref.~\cite{DeWolfe:2007hd},
 the main point to notice is that bulk contributions to the angular potential
 give rise to suppressed by powers of the warp factor. Then, modifications of 
 the angular potential due to bulk effects are generally subleading with 
 respect to the potential produced by moduli stabilization effects on the throat.
 On the other hand, bulk effects could play a role in the infrared region where 
 the angular instabilities develop, and where the overall force on the angular 
 modes due to moduli stabilization effects vanish. But bulk contributions
 are subdominant away from this particular region, being suppressed by the warp 
 factor. Consequently, we expect that their net effect is at most to induce a tiny
 shift of the position at which the instability occurs along the radial direction.

\section{Explicit multi-field potential from warped conifold}
\label{sec:multifield}

In our previous analysis, we mostly focused on studying the homogeneous dynamics of the inflaton field along the extremal trajectories. On the other hand, since towards the later stages of slow-roll inflation an instability develops in the angular directions, in this region the masses of the corresponding fields become small, comparable to that of the radial direction, and eventually tachyonic. To follow the dynamics from this stage onwards, we need to consider in more detail the properties of the angular fields, significantly deviating from their extremal values. With this purpose, in this section we study more closely the potential governing the angular fields that develop an instability towards the end of inflation, as well as the corresponding kinetic terms. This analysis represents the necessary first step for discussing in detail the field evolution in this system at the homogenous level as well as the dynamics of fluctuations. Simple considerations, based on the properties of the potential, allow us to obtain important information about the instability process. In particular, we show that the instability connects one of the extremal trajectories followed along during inflation to another, e.g. non-delta-flat trajectory to delta-flat or non-delta-flat one. This is studied in Section~\ref{genpot3fc} where we also discuss the fact that the actual dynamics of the system is characterized by non-standard kinetic terms which contain cross terms involving derivatives of the fields.

\subsection{Alternative parametrization of the angular fields}

The fields $P$, $Q$ and $R$ that we used in Sections~{\ref{sec:angstaban}} and {\ref{sec:numeric}} are given by combinations of the angular coordinates $\alpha_i$, $\beta_i$ and $\gamma_i$ in the warped conifold,  using the so-called Euler-Rodriguez parametrization discussed in Appendix~{\ref{app:angular}}. Recall that, in Section~{\ref{sec:numeric}}, we found that the angular field $P$ remains heavier than $Q$ and $R$ during the most interesting phase of inflation, and does not take part in the instability process\footnote{Notice that this fact holds only for the specific inflationary model we have considered. There could be other inflationary trajectories, with different choices of the parameters, for which the behavior of $P$ is different.}. Thus we set it at its extremal value $P=0$ in the following discussion. From its definition (\ref{DefP}), this gives $\alpha_1+\alpha_2=0$. We will restrict ourselves to the case $\alpha_1=\alpha_2=0$, which greatly simplifies the following calculations. We then concentrate on the remaining angular modes $Q$ and $R$ in the previous discussion. We have seen in the previous section that the masses of these modes become tachyonic at almost the same point along the inflationary region, see Fig.~\ref{Fig:masses}. For this reason we are interested in determining the complete potential that describes these fields together, as well as the radial direction. As can be read from (\ref{DefQ}) and (\ref{DefR}), they are actually formed by a combinations of two pairs of angular variables. It turns out that, while for small perturbations around the extremal trajectories the combinations of $\beta_1$, $\beta_2$, $\gamma_1$ and $\gamma_2$  giving $Q$ and $R$ are the most convenient to deal with, this is no longer true when describing arbitrary displacements from the extremal points. To describe this last case, it is useful to define another basis of fields.

From (\ref{forforW}) and (\ref{expextle}), we can easily find that starting from a point along the extremal trajectories, $w_1$ and $w_2$ are given by
\begin{align}
\label{w1_angular}
w_1 = & w_1^{(0)}\cos|\chi_1|\cos|\chi_2| - w_2^{(0)}\sin|\chi_1|\sin|\chi_2| \left( \frac{\gamma_1}{|\chi_1|} - i\frac{\beta_1}{|\chi_1|} \right) \left( \frac{\gamma_2}{|\chi_2|} - i\frac{\beta_2}{|\chi_2|} \right) \, ,
\\
\label{w2_angular}
w_2 = & w_2^{(0)}\cos|\chi_1|\cos|\chi_2| + w_1^{(0)}\sin|\chi_1|\sin|\chi_2| \left( \frac{\gamma_1}{|\chi_1|} + i\frac{\beta_1}{|\chi_1|} \right) \left( \frac{\gamma_2}{|\chi_2|} + i\frac{\beta_2}{|\chi_2|} \right) \, ,
\end{align}
with $\chi_i$ given by (\ref{def:chi}) as $\chi_i^2\,=\,\beta_i^2+\gamma_i^2$. The actual range of the quantities $\beta_i$ and $\gamma_i$ depends on the extremal trajectory one considers, as discussed in Appendix~\ref{app:versus} (see also the discussion in the next section), but we can take them positive. Along with $\chi_1$ and $\chi_2$, it is natural to define a field $\xi$ as
\begin{align}
\xi_1+\xi_2 = & 2\xi \, ,
\\
\label{defxi1}
\xi_1 = & \cos^{-1} \left( \frac{\gamma_1}{|\chi_1|} \right) = \sin^{-1} \left( \frac{\beta_1}{|\chi_1|} \right) \, ,
\\
\label{defxi2}
\xi_2 = & \cos^{-1} \left( \frac{\gamma_2}{|\chi_2|} \right) = \sin^{-1} \left( \frac{\beta_2}{|\chi_2|} \right) \, .
\end{align}
It is convenient to consider $\xi$ rather than $\xi_1$ and $\xi_2$ separately since, as can be read from (\ref{w1_angular}) and (\ref{w2_angular}), only the combination $\xi_1+\xi_2$ appears in the potential. With these new fields, and using (\ref{wextTraj}) and (\ref{wextTraj2}), we can write $w_1$ and $w_2$ as
\begin{equation}\label{dfw1}
w_1 = \left\{
 \begin{split}
 & \pm \frac{\epsilon}{\sqrt{2}} e^{\tau/2} L_-^- \pm i\frac{\epsilon}{\sqrt{2}}e^{-\tau/2}\sin|\chi_1|\sin|\chi_2|\sin(2\xi) \, , \qquad \text{(non-delta-flat)}
 \\
 & \pm \frac{\epsilon}{\sqrt{2}} e^{-\tau/2} L_-^+ \pm i\frac{\epsilon}{\sqrt{2}}e^{\tau/2}\sin|\chi_1|\sin|\chi_2|\sin(2\xi) \, , \qquad \text{(delta-flat)}
\end{split}\right.
\end{equation}
\begin{equation}\label{dfw2}
w_2 = \left\{
 \begin{split}
 & \mp\frac{\epsilon}{\sqrt{2}}e^{-\tau/2}L_-^+ \pm i\frac{\epsilon}{\sqrt{2}}e^{\tau/2}\sin|\chi_1|\sin|\chi_2|\sin(2\xi) \, , \qquad \text{(non-delta-flat)}
 \\
 & \mp\frac{\epsilon}{\sqrt{2}}e^{\tau/2}L_-^- \pm i\frac{\epsilon}{\sqrt{2}}e^{-\tau/2}\sin|\chi_1|\sin|\chi_2|\sin(2\xi) \, , \qquad \text{(delta-flat)}
 \end{split}\right.
\end{equation}
where we have defined
\begin{align}
L_+^\pm \equiv & \cos|\chi_1|\cos|\chi_2| + \exp\left( \pm \tau \right)
\sin|\chi_1|\sin|\chi_2|\cos(2\xi) \, ,
\\
L_-^\pm \equiv & \cos|\chi_1|\cos|\chi_2| - \exp\left( \pm \tau \right)
\sin|\chi_1|\sin|\chi_2|\cos(2\xi) \, .
\end{align}
In the limit of vanishing $\chi_1$, $\chi_2$ and $\xi$, we recover (\ref{wextTraj}) and (\ref{wextTraj2}). Note that these expressions have an interesting property: starting from one extremal trajectory, we can move to another by continuously varying $\chi_1$, $\chi_2$ and $\xi$: for example, as will be demonstrated momentarily, from the lower branch of the non-delta-flat extremal trajectory obtained by $\chi_1=\chi_2=\xi=0$, we can obtain both branches of the delta-flat one by taking $|\chi_1|=|\chi_2|=\pi/2$ and $\xi=0$ or $\xi=\pi/2$. We will discuss some consequences of this fact in what follows.

\subsection{Potential and kinetic terms}
\label{genpot3fc}

Having found $w_1$ and $w_2$ as (\ref{dfw1}) and (\ref{dfw2}), we can straightforwardly calculate the scalar potential. After some computations, we can find that the form (\ref{VF_before_angular_stab}) does not change, but the functions $\calG$,  $\calC$, and $\calD$  become different. The results are
\begin{align}
\calG = &
 \begin{cases}
  \left( 1 \mp \alpha e^{\tau/2}L_-^- \right)^2 + \alpha^2e^{-\tau} \left[ \sin|\chi_1|\sin|\chi_2|\sin(2\xi) \right]^2 \, , & \text{(non-delta-flat)}
  \\
  \left( 1 \mp \alpha e^{-\tau/2}L_-^+ \right)^2 + \alpha^2e^\tau \left[ \sin|\chi_1|\sin|\chi_2|\sin(2\xi) \right]^2 \, , & \text{(delta-flat)}
 \end{cases}
\\
\calC = &
 \begin{cases}
  2\tanh\tau \left\{ \pm\alpha e^{\tau/2}L_+^- - \alpha^2e^\tau L_-^-L_+^- + \alpha^2 e^{-\tau} \left[ \sin|\chi_1|\sin|\chi_2|\sin(2\xi) \right]^2 \right\} \, , & \text{(non-delta-flat)}
  \\
  2\tanh\tau \left\{ \mp\alpha e^{-\tau/2}L_+^+ + \alpha^2e^{-\tau}L_-^+L_+^+ - \alpha^2e^\tau \left[ \sin|\chi_1|\sin|\chi_2|\sin(2\xi) \right]^2 \right\} \, , & \text{(delta-flat)}
 \end{cases}
\\
\calD = &
 \begin{cases}
  1 - \sech\tau \left\{ \cfrac{e^{-\tau}}{2}{L_-^+}^2 + \cfrac{e^\tau}{2}\left[ \sin|\chi_1|\sin|\chi_2|\sin(2\xi) \right]^2 \right\} + \left( \cfrac{k''}{k'} - \coth\tau \right)
  \\
  \times \left\{ \tanh\tau + 2\csch(2\tau) \left[ (\cos|\chi_1|\cos|\chi_2|)^2 + (\sin|\chi_1|\sin|\chi_2|)^2 + \cfrac{1}{2}\cosh\tau\sin(2|\chi_1|)\sin(2|\chi_2|)\cos(2\xi) \right] \right.
  \\
  \left. \hspace{0.5cm} - \sech\tau \left( \cfrac{e^\tau}{2}{L_-^-}^2 + \cfrac{e^{-\tau}}{2}{L_-^+}^2 + \cosh\tau \left[ \sin|\chi_1|\sin|\chi_2|\sin(2\xi) \right]^2 \right) \right\} \, , \quad \text{(non-delta-flat)}
  \\
  1 - \sech\tau \left\{ \cfrac{e^{\tau}}{2}{L_-^-}^2 + \cfrac{e^{-\tau}}{2}\left[ \sin|\chi_1|\sin|\chi_2|\sin(2\xi) \right]^2 \right\} + \left( \cfrac{k''}{k'} - \coth\tau \right)
  \\
  \times \left\{ \tanh\tau + 2\csch(2\tau) \left[ (\cos|\chi_1|\cos|\chi_2|)^2 + (\sin|\chi_1|\sin|\chi_2|)^2 + \cfrac{1}{2}\cosh\tau\sin(2|\chi_1|)\sin(2|\chi_2|)\cos(2\xi) \right] \right.
  \\
  \left. \hspace{0.5cm} - \sech\tau \left( \cfrac{e^{-\tau}}{2}{L_-^+}^2 + \cfrac{e^{\tau}}{2}{L_-^-}^2 + \cosh\tau \left[ \sin|\chi_1|\sin|\chi_2|\sin(2\xi) \right]^2 \right) \right\} \, . \quad \text{(delta-flat)}
 \end{cases}
\end{align}
Plugging these expressions into (\ref{VF_before_angular_stab}), we can obtain the full potential including angular directions. This is the complete form of the potential that is needed in order to study the dynamics of the angular and radial directions in the instability region. While in Section~\ref{sec:numeric} we focussed on the lower branch of the non-delta flat trajectory, the previous formulae allow us to also obtain the potential in all other cases. In the limit of small $\chi_i$, or equivalently of small $Q$ and $R$, this potential coincides with the one we used in Appendix~\ref{app:angular} to determine the eigenvalues $X_Q$ and $X_R$.

The potential is periodic along  the angular directions $\chi_1, \chi_2$ and $\xi$, as it is clear given our expressions for $w_1$ and $w_2$. There are ``{\em extremal points}'', defined as the positions for which derivatives of the potential along $\chi_i$ and $\xi$ vanish. These extremal points coincide with the extremal points along the directions $Q$ and $R$: namely they provide the same non-delta-flat and delta-flat trajectories determined in Appendix~\ref{app:angular}. They are shown in Table~\ref{Table:extremal}, but others that are periodically identifiable with these ones are also extremal points.
\begin{table}[h]
 \begin{center}
  \begin{tabular}{c|c*{4}{c}}
   \hline\hline
              & NDF, L & NDF, U  & NDF, U  & DF, U   & DF, L
   \\
   \hline
   $|\chi_1|$ & 0      & 0       & $\pi$   & $\pi/2$ & $\pi/2$
   \\
   $|\chi_2|$ & 0      & $\pi$   & 0       & $\pi/2$ & $\pi/2$
   \\
   $\xi$      & 0      & $\pi/2$ & $\pi/2$ & 0       & $\pi/2$
   \\
   \hline\hline
  \end{tabular}
 \end{center}
 \caption{The values of $\chi_1$, $\chi_2$ and $\xi$ for extremal points. ``NDF'' and ``DF'' stand for non-delta-flat and delta-flat, and ``U'' and ``L'' for upper and lower branches respectively. For example, the first column indicates the lower branch in the non-delta flat trajectory, the case on which we focussed in Section~4.}
 \label{Table:extremal}
\end{table}

\begin{figure}[h]
 \begin{center}
  \includegraphics[width = 15cm]{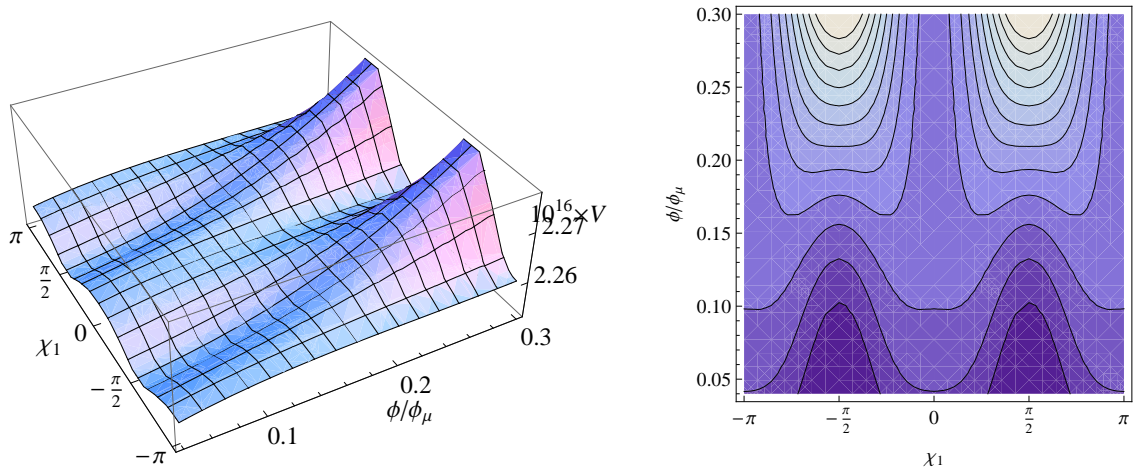}\vspace{-1.5em}
 \end{center}
 \caption{The multi-field potential as a function of the radial direction and a particular choice of angular coordinates $\chi_1=\chi_2$ and $\xi=0$. We can see that other stable trajectories exist at $|\chi_1| = |\chi_2| = \pi/2$ next to the unstable one at $\chi_1=\chi_2=0$. The potential is periodic along the direction $\chi_1$, with period $\pi$. This is consistent with our observation that $|\chi_i|\in[0,\pi/2]$: see the last part of Appendix~B.3.
 We used the other parameters the same as in Table~1.}
 \label{fig:multifield}
\end{figure}

These observations are sufficient to extract important qualitative information about the outcome of the instability process. For definiteness, let us consider the explicit example we studied in the previous section, which corresponds to the first column of Table~\ref{Table:extremal}. For the choice of parameters
 summarised in Table~\ref{Table:choice}, all of the extremal points 
 have at least one unstable angular mode except for the upper branch of 
 the delta-flat direction ($|\chi_1|=|\chi_2|=\pi/2$), which corresponds 
 to the true vacuum of the D3-brane along the angular directions. Until 
 the point at which instability develops, all the angular modes are at 
 the origin. Nearby the instability region, the non-delta-flat trajectory 
 becomes a local maximum (or saddle point) for the angular modes. 
The angular masses become light and then tachyonic, and the fields $\chi_i$ and $\xi$ start to roll down the potential. In this case, the range of the variables are found to be $\xi\in[0,4\pi]$ and $\chi_i \in[-\pi/2,\pi/2]$ (or equivalently $|\chi_i|\in[0,\pi/2]$), as discussed in the last part of Appendix~\ref{app:versus}. After a possibly complicated dynamics, they reach another extremal point of lower energy, at which all the squares of the mass eigenvalues are positive with our choice of parameters. This corresponds to the upper branch of the delta-flat trajectory, the fourth column of Table~\ref{Table:extremal}. The actual result depends on the detailed dynamics of the system. On the other hand, our knowledge of the potential is sufficient to show that the angular field dynamics, starting at the point of instability, connects the initial non-delta-flat with one of the delta-flat trajectories. As a representative example, we plot in Fig.~\ref{fig:multifield} the potential depending on both radial and angular directions, taking for definiteness $\chi_1=\chi_2$ and keeping $\xi=0$. The plot clearly exhibits an instability along the non-delta-flat direction ($\chi_1=\chi_2=0$), and we can more easily see that the potential connects the false vacuum corresponding to the lower branch of the non-delta-flat trajectory with the true vacuum corresponding to the upper branch of the delta-flat trajectory ($|\chi_1|=|\chi_2|=\pi/2$) in the angular directions.

As we have seen from the previous arguments, the angular instability connects different extremal trajectories in the warped deformed conifold. Depending on initial conditions, after the field falls from one trajectory to the other, a non-trivial dynamics continues along the radial and, possibly, along the  angular directions. This implies that, at the later stages of slow-roll inflation, the inflaton field does not generically fall into a global minimum, but into other extremal points that further evolve due to non-trivial motion in the radial direction. In this setup, consequently, it is not possible to obtain a quantitative understanding of the process of formation and evolution of topological defects at the end of inflation, based on the symmetries of the potential. Instead, a detailed numerical analysis is necessary, which takes into account of the post-inflationary evolution along the radial direction.

The analysis of the actual dynamics that develops at the instability is further complicated by the non-standard form of the kinetic terms characterizing the fields. In order to obtain them, one can follow an identical procedure to the one discussed in Section~\ref{sec:angstaban} and Appendix~\ref{app:angular}. After straightforward calculations, we can find that the form of the kinetic term is given by (\ref{D3KE}) where the pull-back $d\hat{s}_6^2$ given by (\ref{eq:DeformMetric}), with
\begin{align}
g^{(1)} = & \sin(2\chi_1)\,d \xi_1 + \cos(2\xi)\sin(2\chi_2)\,d \xi_2 + 2 \sin(2\xi)\,d \chi_2 \, ,
\\
g^{(2)} = & 2 d \chi_1 + \sin(2\xi)\sin(2\chi_2)\,d \xi_2 - 2\cos(2\xi)\, d \chi_2 \, ,
\\
g^{(3)} = & \sin(2\chi_1)\,d \xi_1 - \cos(2\xi)\sin(2\chi_2)\,d \xi_2 - 2\sin(2\xi)\,d \chi_2 \, ,
\\
g^{(4)} = & 2 d \chi_1 - \sin(2\xi)\sin(2\chi_2)\,d \xi_2 + 2\cos(2\xi) \, d \chi_2 \, ,
\\
g^{(5)} = & \left[1-\cos(2\chi_1)\right]\,d \xi_1 + \left[1-\cos(2\chi_2)\right]\,d \xi_2 \, .
\end{align}
The kinetic terms are then characterized by the cross terms involving derivatives of the fields. These cross terms have very small coefficients when considering small deviations around the extremal trajectory. But they play an important role when large angular displacements are considered, which is expected during the instability process. The range of the angular fields, in the previous equations, depends on the initial trajectory one is considering, as discussed in the first part of this section.

In this section, we have provided all the necessary tools that are needed in order  to study the dynamics of the instability process at later stages of inflation. We have obtained important qualitative information about the field dynamics and the instability process. These ingredients are the starting point to perform a more detailed analysis 
 of the field evolution after the instability develops.
 For example, it would be interesting to study how the 
 annihilation of D3-$\overline{\rm D3}$ branes occurs, 
 the role of the angular directions in this process,
 and whether moduli stabilization induced by the presence
 of D7-branes can have some effects in this phenomenon.
 Indeed, in Ref.~\cite{DeWolfe:2007hd} it has been shown 
 that  near the tip of the throat the force between brane 
 and antibrane can acquire significant angular dependence.
 Another important issue is
to investigate possible formation and evolution of topological defects after inflation ends. We hope to return to discuss these issues elsewhere.

\section{Future Directions}
\label{sec:future}

In this work, we have considered a model of D3-brane inflation in a warped throat. Different forces act on the D3-brane, due to the presence of an anti D3-brane at the tip of the throat, and of a Ouyang-embedded D7-brane in the warped deformed conifold. We have  shown that these forces  compensate each other, and can obtain 60 $e$-foldings of slow-roll inflation around an inflection point of the potential along the radial direction. At later stages of single field slow-roll inflation, the angular directions develop instabilities after which more complicated dynamics follows. Our work provides a new explicit example for obtaining  slow-roll inflation in warped throats, other than using Kuperstein embedded D7-brane~\cite{Baumann:2007ah,Chen:2008ai}. Moreover, it is an example where angular directions play an active role for determining the inflationary dynamics while the model is fully under control from the supergravity point of view and the complete potential for the relevant modes is known.

There are various topics, motivated by our study, that deserve further analysis. We have shown that the masses of the angular modes remain comparable to the Hubble scale during most of the inflationary period, becoming light towards the later stages of slow-roll inflation. The fact that the angular masses are not  much larger than the Hubble scale can lead to a situation in which the slow-roll motion continues along
the radial direction for few $e$-foldings, even after the development of the instability in the angular directions. When the slow-roll conditions are violated during the instability process, however, the DBI nature of the D-brane action shows up, since the brane velocity increases. This fact can also play an important role in determining the detailed dynamics governing the instability. It would be interesting to study the consequences of these features for the spectrum of the curvature perturbation, taking into account the presence of the non-canonical kinetic terms for the fields involved, and the particular form of the potential.

Another important direction is to study the formation and stability of topological defects produced after inflation. As we have argued, the instability process connects different extremal trajectories. Consequently, after the onset of instability, the system does not reach a global minimum, and further motion of the brane along the radial direction takes place. The study of implications of this brane motion, for a possible development of topological defects, needs a detailed numerical investigation. Finally, it would  also be interesting to understand whether the energy stored in the inflaton field can be transmitted into the matter sector, while the angular fields roll off their false minima. We hope to return to analyse these issues in the near future.

\acknowledgments

We would like to thank  Ruth Gregory, Liam McAllister, Peter Ouyang, Gary Shiu, and Ivonne Zavala for very useful discussions.
HYC would like to thank PCTS, Princeton University, where this work was initiated. He is supported in part by NSF CAREER Award No. PHY-0348093, DOE grant DE-FG-02-95ER40896, a Research Innovation Award and a Cottrell Scholar Award from Research Corporation, and a Vilas Associate Award from the University of Wisconsin.
JG is grateful to the Yukawa Institute for Theoretical Physics, Kyoto University for hospitality during the long-term workshop ``Gravity and Cosmology 2010 (GC2010)'' (YITP-T-10-01) and the YKIS symposium ``Cosmology -- The Next Generation --'' (YKIS2010) where this work was being finished.
He is partly supported by a VIDI and a VICI Innovative Research Incentive Grant from the Netherlands Organisation for Scientific Research (NWO).
KK would like to thank the Yukawa Institute for Theoretical Physics, Kyoto University and the Royal Society for two workshops, ``Non-linear cosmological perturbations'' (YITP-W-09-01) and ``The non-Gaussian universe'' (YITP-T-09-05) where he is benefitted from many stimulating discussions.
He is supported by European Research Coucil, Research Councils UK and STFC.

\newpage

\appendix
\section*{Appendices}

\section{Properties of warped deformed conifold}
\label{app:conifold}

\subsection{Parametrization of the angular coordinates}

Here we collect some facts concerning the various coordinates parameterizing the deformed conifold following Ref.~\cite{Candelas:1989js}. It can be defined as a submanifold ${\mathbb{C}}^4$ via the complex embedding equation
\begin{equation}\label{zDefeqnconifold}
\sum_{\alpha=1}^4 (z^\alpha)^2 = \epsilon^2 \, ,
\end{equation}
where $\{z^\alpha\}$ are complex coordinates in ${\mathbb{C}}^4$. They can be expressed in terms of a $2\times 2$ matrix as
\begin{equation}
\label{Wdef}
W = \begin{pmatrix}
-w^3 \; w^2 \cr -w^1 \; w^4
\end{pmatrix}
= \frac{1}{\sqrt{2}}
\begin{pmatrix}
z^3 + i z^4 \;\;\;\; z^1 - i z^2 \cr z^1 + i
z^2 \;\; -z^3 + i z^4
\end{pmatrix} \, .
\end{equation}
The deformed conifold relation (\ref{zDefeqnconifold}) can then be written as
\begin{equation}\label{Detwcoord}
{\rm det}\ W = w^1 w^2 - w^3 w^4=- \frac{1}{2} \epsilon^2 \, .
\end{equation}
The complex embedding coordinates of deformed conifold $\{z^1,z^2,z^3,z^4\}$ can also be expressed in terms of the real global coordinates $\{\tau\in{\mathbb{R}\,,~\psi\in[0,4\pi]\,,~\theta_{1,2}\in[0,\pi]\,,~\phi_{1,2}\in[0,2\pi]}\}$, $\Xi=\tau+i\psi$ as
\begin{align}
\label{FullDefConifoldcoord1}
z^1 = & \epsilon \left[ \cosh \left(\frac{\Xi}{2}\right) \cos \left(\frac{\theta_1+\theta_2}{2}\right) \cos\left(\frac{\phi_1+\phi_2}{2}\right) + i \sinh \left(\frac{\Xi}{2}\right) \cos \left(\frac{\theta_1-\theta_2}{2}\right) \sin \left(\frac{\phi_1+\phi_2}{2}\right) \right] \, ,
\\
\label{FullDefConifoldcoord2}
z^2 = & \epsilon \left[ -\cosh \left(\frac{\Xi}{2}\right) \cos \left(\frac{\theta_1+\theta_2}{2}\right) \sin \left(\frac{\phi_1+\phi_2}{2}\right) + i \sinh \left(\frac{\Xi}{2}\right) \cos \left(\frac{\theta_1-\theta_2}{2}\right) \cos \left(\frac{\phi_1+\phi_2}{2}\right)\right] \, ,
\\
\label{FullDefConifoldcoord3}
z^3 = & \epsilon \left[ -\cosh \left(\frac{\Xi}{2}\right) \sin \left(\frac{\theta_1+\theta_2}{2}\right) \cos \left(\frac{\phi_1-\phi_2}{2}\right) + i \sinh \left(\frac{\Xi}{2}\right) \sin \left(\frac{\theta_1-\theta_2}{2}\right) \sin \left(\frac{\phi_1-\phi_2}{2}\right)\right] \, ,
\\
\label{FullDefConifoldcoord4}
z^4 = & \epsilon \left[ -\cosh \left(\frac{\Xi}{2}\right) \sin \left(\frac{\theta_1+\theta_2}{2}\right) \sin \left(\frac{\phi_1-\phi_2}{2}\right) - i \sinh \left(\frac{\Xi}{2}\right) \sin \left(\frac{\theta_1-\theta_2}{2}\right) \cos\left(\frac{\phi_1-\phi_2}{2}\right)\right] \, .
\end{align}
One can readily verify that the constraints (\ref{zDefeqnconifold}) and (\ref{Detwcoord}) are satisfied. We can also encode the coordinates (\ref{FullDefConifoldcoord1})-(\ref{FullDefConifoldcoord4}) compactly in the matrix form as
\begin{equation}\label{expMatrixcoord}
W \, = \, L_c \, W_s \, R_c^{\dagger} \, ,
\end{equation}
where the $2\times 2$ matrices $W_s$, $L_c$ and $R_c$ are given by
\begin{align}
\label{sppoint}
W_s = & \left(\begin{array}{cc}
  0 & \cfrac{\epsilon}{\sqrt{2}}e^{\tau/2} \\
  \cfrac{\epsilon}{\sqrt{2}}e^{-\tau/2} & 0 \\
\end{array}\right) \, ,
\\
\label{cmatricesdef}
L_c = & \left(
 \begin{array}{cc}
  \cos\left(\cfrac{\theta_1}{2}\right) \exp \left[ \cfrac{i}{2}(\psi_1+\phi_1) \right]
  &\hskip0.5cm -\sin\left(\cfrac{\theta_1}{2}\right) \exp \left[ -\cfrac{i}{2}(\psi_1-\phi_1) \right] \\
  \sin\left(\cfrac{\theta_1}{2}\right) \exp \left[ \cfrac{i}{2}(\psi_1-\phi_1) \right]
  & \hskip0.5cm \cos\left(\cfrac{\theta_1}{2}\right) \exp \left[ -\cfrac{i}{2}(\psi_1+\phi_1) \right] \\
 \end{array}
\right) \, ,
\end{align}
and similarly for $R_c$ with the subscript 1 replaced by 2. $L_c$ and $R_c$ are two independent $SU(2)$ group elements, which generate the entire deformed conifold by acting on the special point $W_s$. Notice that the coordinates $\psi_1$ and $\psi_2$ only appear in the combination $\psi=\psi_1+\psi_2$.

\subsection{The metric of the warped deformed conifold}

It is convenient to define the following basis of one forms~\cite{Klebanov:2000hb,Minasian:1999tt},
\begin{align}\label{Defgvier}
g^{(1)} \equiv & \frac{-\sin\theta_1 d\phi_1-(\cos \psi \sin \theta_2
d\phi_2 - \sin\psi d\theta_2)}{\sqrt{2}} \, ,
\\
g^{(2)} \equiv & \frac{d\theta_1-(\sin\psi \sin \theta_2 d\phi_2
+ \cos \psi d\theta_2)}{\sqrt{2}} \, ,
\\
g^{(3)} \equiv & \frac{-\sin\theta_1 d\phi_1
+(\cos \psi \sin \theta_2 d\phi_2 - \sin\psi d\theta_2)}{\sqrt{2}} \, ,
\\
g^{(4)} \equiv & \frac{d\theta_1+(\sin\psi \sin \theta_2 d\phi_2
+ \cos \psi d\theta_2)}{\sqrt{2}} \, ,
\\
g^{(5)} \equiv & d\psi + \cos \theta_1 d\phi_1 + \cos\theta_2 d\phi_2 \, .
\end{align}
The explicit metric of the six dimensional internal part of the geometry, the deformed conifold, can
be expressed as
\begin{align}\label{eq:DeformMetric}
ds_6^2 = & \frac{\epsilon^{4/3}}{2} K(\tau) \left\{ \frac{1}{3K^3(\tau)} \left[ d\tau^2 + \left( g^{(5)} \right)^2 \right] + \cosh^2 \left( \frac{\tau}{2} \right) \left[ \left( g^{(3)} \right)^2 + \left( g^{(4)} \right)^2 \right] \right.
\nonumber\\
& \left. \hspace{1.8cm} + \sinh^2 \left( \frac{\tau}{2} \right) \left[ \left( g^{(1)} \right)^2 + \left( g^{(2)} \right)^2 \right] \right\} \, ,
\end{align}
where
\begin{equation}\label{DefKtau}
K(\tau) \equiv \frac{\left[ \sinh(2\tau)-2\tau \right]^{1/3}}{2^{1/3}\sinh\tau} \, .
\end{equation}
We can also write the full ten dimensional warped metric as
\begin{equation}\label{Def10Dwarpmetric}
ds_{10}^2 = e^{2A(y)} \eta_{\mu\nu} dx^\mu dx^\nu + e^{-2A(y)} ds_6^2 \, .
\end{equation}

The warping is obtained by turning $M$ units of $F_3$ flux through the $A$-cycle of the deformed conifold, and $-K$ units of $H_3$ flux through the dual $B$-cycle, when we put a ultraviolet cutoff of the warped throat at $e^{-4A} \sim 1$. The warp factor is given by the expression~\cite{Klebanov:2000hb}
\begin{equation}\label{Defwarpfactor}
e^{-4A(\tau)} = 2^{2/3} (g_s M \alpha')^2 \epsilon^{-8/3} I(\tau) \, ,
\end{equation}
where
\begin{equation}\label{DefItau}
I(\tau) \equiv \int_{\tau}^\infty dx \frac{x \coth x -1}{\sinh^2x} \left[ \sinh(2x) - 2x \right]^{1/3} \, .
\end{equation}
The internal, deformed conifold metric (\ref{eq:DeformMetric}) can be obtained from the so-called ``little'' K\"ahler potential $k(z^\alpha,\bar{z}^{\bar{\beta}})$ as
\begin{equation}
\tilde{g}_{\alpha\bar{\beta}} = \partial_\alpha \partial_{\bar{\beta}} k \, .
\end{equation}
As the angular directions of the warped deformed conifold are isometries, they do not appear explicitly in the little K\"ahler potential. It thus only depends on the radial coordinate $\tau$. An expression for
$k(\tau)$ is~\cite{Candelas:1989js}
\begin{equation}\label{DefKahdefcon}
k(\tau) = \frac{\epsilon^{4/3}}{2^{1/3}} \int_0^\tau d\tau' \left[ \sinh(2\tau')-2\tau' \right]^{1/3} \, ,
\end{equation}
where without loss of generality we set the integration constant to zero. For the calculation of the $F$-term scalar potential, the unwarped inversed K\"ahler metric in terms of the holomorphic embedding coordinates $\{w^{\alpha}\}$ is also needed and is given by
\begin{align}
\label{invmetric}
k^{\bar{\alpha}\beta} = & \frac{r^3}{k''} \left[ R^{\bar{\alpha}\beta} + \coth\tau \left( \frac{k''}{k'}
- \coth\tau \right) L^{\bar{\alpha}\beta} \right] \, ,
\\
\label{Rij}
R^{\bar{\alpha}\beta} = & \delta^{\bar{\alpha}\beta} - \frac{c^{\alpha}_{\alpha'}c^{\beta}_{\beta'}w^{\alpha'}\overline{w^{\beta'}}}{r^3} \, ,
\\
\label{Lij}
L^{\bar{\alpha}\beta} = & \left( 1 - \frac{\epsilon^4}{r^6} \right) \delta^{\bar{\alpha}\beta} +
\frac{\epsilon^2}{r^3} \frac{c^{\beta}_{\beta'} w^\alpha w^{\beta'} + c^{\beta}_{\beta'}\overline{w^{\alpha}}\overline{w^{\beta'}}}{r^3} - \frac{w^\alpha\overline{w^\beta}
+ c_{\alpha}^{\alpha'}c_{\beta}^{\beta'}\overline{w^{\alpha'}}w^{\beta'}}{r^3} \, .
\end{align}
Here $c^{\alpha}_{\alpha'}$ is a $4\times 4$ matrix whose only non-zero entries are $c^{1}_2=c^2_1=-1$ and $c^3_4=c^4_3=1$. Notice that the metric (\ref{invmetric}) is indeed six dimensional as the deformed conifold constraint $w_1 w_2-w_3 w_4=-\epsilon^2/2$ allows us to substitute away e.g. $w_4$ and $\overline{w_4}$.

To recover the familiar singular conifold limit, we can define the radial coordinate $r$ as
\begin{equation}\label{Defr}
r^3 = \sum_{\alpha=1}^4 |z^{\alpha}|^2=\epsilon^2\cosh\tau \, .
\end{equation}
In the asymptotic limit $e^{\tau}\to \infty$, the metric (\ref{eq:DeformMetric}) reduces to usual conical form
\begin{equation}\label{SingConMetric}
ds^6\to \frac{3}{2}\left(dr^2+r^2 ds^2_{T^{1,1}}\right) \, ,
\end{equation}
where the metric $ds^2_{T^{1,1}}$ for the five dimensional base $T^{1,1}$ can be easily deduced from (\ref{eq:DeformMetric}). Furthermore the warp factor $e^{-4A}$ (\ref{Defwarpfactor}) acquires an AdS form
\begin{equation}\label{asymwarpfactor}
e^{-4A(r)}\approx \frac{3\pi g_s N(\alpha')^2}{r^4} \, ,
\end{equation}
where $N$ is the quantized five form flux.

\section{Extremal trajectory and angular mass matrix}
\label{app:angular}

In Section~\ref{sec:conifold}, we derived the complete potential controlling the dynamics of the D3-brane along the throat. It explicitly depends on the D3-brane coordinates, in particular on the angular directions. In this section, we determine extremal trajectories of the D3-brane that extremize the potential along the angular directions. These extremal trajectories, along which the brane moves only in the radial direction, correspond to stable points in the moduli space of angular directions, at least for sufficiently large radius $r^3 = \epsilon^2 \cosh{\tau}$.  For small values of $r$, instead, extremal trajectories become unstable for some of the angular directions. The consequences of this fact for inflation is discussed in Sections~\ref{sec:numeric} and \ref{sec:multifield}.

We denote the five angular coordinates of deformed conifold by $\{\Psi_i\}$. We then look for extremal trajectories that satisfy
\begin{equation}\label{AngStabCond}
\frac{\partial{V}}{\partial\Psi_i} = 0 \, .
\end{equation}
The chain rule ensures that finding points satisfying (\ref{AngStabCond}) is equivalent to identifying  points such that
\begin{equation}\label{coorStabcond}
\frac{\partial{ \left( w_1+\overline{w_1} \right) }}{\partial\Psi_i} = \frac{\partial{|w_1|^2}}{\partial\Psi_i} = \frac{\partial{ \left( w_2+\overline{w_2}
\right) }}{\partial\Psi_i} = \frac{\partial{|w_2|^2}}{\partial\Psi_i} =
\frac{\partial{ \left( w_1w_2+\overline{w_1}\overline{w_2} \right)
}}{\partial\Psi_i} = 0 \, .
\end{equation}
This is because, as we can read from (\ref{VF_before_angular_stab}), the $F$-term scalar potential $V_F$ only depends on these combinations of coordinates of the deformed conifold. Thus, our next task is to relate the brane coordinates $w_\alpha$ to the angular coordinates $\Psi_i$.

To parametrize the angular coordinate $\Psi_i$, we can rewrite the matrix $W$ as~\cite{Baumann:2007ah}
\begin{equation}\label{forforW}
W \equiv L_e W_0 R_e^\dag \, .
\end{equation}
Here $W_0$ is given by
\begin{equation}\label{W02}
W_0 = \left(
 \begin{array}{cc}
  -w_3^{(0)} & w_2^{(0)} \\
  -w_1^{(0)} & w_4^{(0)} \\
 \end{array}
\right) \, ,
\end{equation}
and denotes a fiducial point, which we consider to be angularly stable. The matrices $L_e$ and $R_e$ are two independent $SU(2)$ group elements. Using the generators of the $SU(2)$, i.e. the Pauli matrices
\begin{align}
\sigma_1 = &
\left(%
\begin{array}{cc}
  0 & 1 \\
  1 & 0 \\
\end{array}%
\right) \, ,
\\
\sigma_2 = &
\left(%
\begin{array}{cc}
  0 & -i \\
  i & 0 \\
\end{array}%
\right) \, ,
\\
\sigma_3 = &
\left(%
\begin{array}{cc}
  1 & 0 \\
  0 & -1 \\
\end{array}%
\right) \, ,
\end{align}
$L_e$ and $R_e$ can then be identified as
\begin{align}
\label{Wmatrix_stable1}
L_e = & e^{iT_1} \, ,
\\
\label{Wmatrix_stable2}
R_e = & e^{iT_2} \, ,
\end{align}
where $T_1$ and $T_2$ are parametrized by two real three-vectors $\left\{ \alpha_i,\beta_i,\gamma_i \right\}$ ($i=1,2$) as
\begin{equation}\label{generator_matrix}
T_i = \beta_i\sigma_1 + \gamma_i\sigma_2 + \alpha_i\sigma_3 = \left(%
\begin{array}{cc}
  \alpha_i & \beta_i-i\gamma_i \\
  \beta_i+i\gamma_i & -\alpha_i \\
\end{array}%
\right) \, .
\end{equation}
Denoting
\begin{equation}\label{def:chi}
\chi_i^2\,\equiv\, \alpha_i^2+\beta_i^2+\gamma_i^2 \, ,
\end{equation}
a simple calculation provides
\begin{equation}\label{expextle}
L_e = \left(%
\begin{array}{cc}
 \cos{|\chi_1|}+i\alpha_1\,\cfrac{\sin{|\chi_1|}}{|\chi_1|} &
 i \left(\beta_1-i\gamma_1\right)\,\cfrac{\sin{|\chi_1|}}{|\chi_1|}
 \\
 i \left(\beta_1+i\gamma_1\right)\,\cfrac{\sin{|\chi_1|}}{|\chi_1|} &
 \cos{|\chi_1|}-i\alpha_1\,\cfrac{\sin{|\chi_1|}}{|\chi_1|}
\end{array}%
\right) \, ,
\end{equation}
and similarly for $R_e$.

The two $SU(2)$ group elements $L_e$ and $R_e$ take a point $W_0$ on deformed conifold to the entire manifold. Notice that here we have adopted different parametrization of $SU(2)$ group element (Euler-Rodriguez) from the one used in (\ref{expMatrixcoord}) (Cayley-Kline), as the former parametrization is particularly suitable to study the properties of extremal trajectories. The parameters $\{\alpha_i\,,\beta_i\,, \gamma_i\}$ ($i=1,2$) of (\ref{expextle}) and $\{\theta_i\,,\phi_i\,,\psi_i\}$ ($i=1,2$) of (\ref{cmatricesdef}) are usually related non-linearly. We will discuss the relation between these two different parametrizations in Appendix~\ref{app:versus}.

\subsection{Linear expansion: Identifying the extremal trajectory}

As we want angular stability near $w_\alpha^{(0)}$ satisfying (\ref{coorStabcond}), expanding (\ref{Wmatrix_stable1}) and (\ref{Wmatrix_stable2}) and only keeping the linear order, then we find
\begin{equation}
W = \left( \openone + iT_1 \right) W_0 \left( \openone - iT_2 \right) = W_0 + i \left( T_1W_0 - W_0T_2 \right) \, .
\end{equation}
The changes in $w_1$ and $w_2$ are then given by
\begin{align}
\label{delta_w1}
\delta{w_1} = & -i(\alpha_1+\alpha_2)w_1^{(0)} + (i\beta_1-\gamma_1)w_3^{(0)} +
(i\beta_2-\gamma_2)w_4^{(0)} \, ,
\\
\label{delta_w2}
\delta{w_2} = & i(\alpha_1+\alpha_2)w_2^{(0)} + (i\beta_1+\gamma_1)w_4^{(0)} +
(i\beta_2+\gamma_2)w_3^{(0)} \, .
\end{align}
Then, it is trivial to find, by demanding the linear variations of the combinations of $w_1$ and $w_2$ which appear in the potential vanish, that $\left\{ w_1^{(0)}, w_2^{(0)} \right\} \in \mathbb{R}$ and $w_3^{(0)} = w_4^{(0)} = 0$.
Thus, the conifold constraint equation now gives
\begin{equation}\label{extremal_deformed_conifold}
w_1^{(0)}w_2^{(0)} = -\frac{\epsilon^2}{2} \, .
\end{equation}
Combining with (\ref{Defr}), which now reads
\begin{equation}\label{extremal_radial_coordinate}
\left|w_1^{(0)}\right|^2 + \left|w_2^{(0)}\right|^2 = \left(w_1^{(0)}\right)^2 + \left(w_2^{(0)}\right)^2 = \epsilon^2\,\cosh{\tau} \, ,
\end{equation}
we determine the following extremal trajectories
\begin{equation}\label{ext_traj1}
w_1^{(0)} =  \pm \frac{\epsilon}{\sqrt{2}} e^{\tau/2} \, , \qquad w_2^{(0)} = \mp \frac{\epsilon}{\sqrt{2}} e^{-\tau/2} \, ,
\end{equation}
and
\begin{equation}\label{ext_traj2}
w_1^{(0)} = \pm \frac{\epsilon}{\sqrt{2}} e^{-\tau/2} \, , \qquad w_2^{(0)} = \mp \frac{\epsilon}{\sqrt{2}} e^{\tau/2} \, .
\end{equation}
In the large radius limit $e^{\tau/2}\to \infty$, the trajectories (\ref{ext_traj1}) and (\ref{ext_traj2}) correspond respectively to the so-called {\em non-delta-flat} and {\em delta-flat} trajectories determined in Ref.~\cite{Baumann:2007ah}. While we present an explicit realization of slow-roll inflation using the non-delta-flat trajectory, we will also discuss the role of the delta-flat trajectory for the multi-field inflationary models.

\subsection{Quadratic expansion: Mass matrix}

We now develop the tools necessary to investigate the stability of the extremal trajectories (\ref{ext_traj1}) and (\ref{ext_traj2}) along the angular directions. To find the mass matrix associated with the angular coordinates, we expand (\ref{Wmatrix_stable1}) and (\ref{Wmatrix_stable2}) up to second order in perturbations. Explicitly, we have
\begin{align}\label{Wexpansion1}
W = & \left( \openone + iT_1 - \frac{1}{2}T_1^2 \right) W_0 \left( \openone - iT_2 - \frac{1}{2}T_2^2 \right)
\nonumber\\
= & W_0 + i \left( T_1W_0 - W_0T_2 \right) + \left[ T_1W_0T_2 - \frac{1}{2} \left( T_1^2W_0 + W_0T_2^2 \right) \right] \, .
\end{align}
Using $w_3^{(0)} = w_4^{(0)} = 0$ along the extremal trajectories, we can explicitly calculate the combinations $\left\{w_1+\overline{w_1}, |w_1|^2, w_2+\overline{w_2},
|w_2|^2, w_1w_2+\overline{w_1}\overline{w_2}\right\}$, and find
\begin{align}
w_1+\overline{w_1} = & 2w_1^{(0)} - \left[ (\alpha_1+\alpha_2)^2 + \beta_1^2 + \beta_2^2 + \gamma_1^2 + \gamma_2^2 \right] w_1^{(0)} - 2 \left( \beta_1\beta_2 - \gamma_1\gamma_2 \right) w_2^{(0)} \, ,
\\
w_2+\overline{w_2} = & 2w_2^{(0)} - \left[ (\alpha_1+\alpha_2)^2 + \beta_1^2 + \beta_2^2 + \gamma_1^2 + \gamma_2^2 \right] w_2^{(0)} - 2 \left( \beta_1\beta_2 - \gamma_1\gamma_2 \right) w_1^{(0)} \, ,
\\
\left|w_1\right|^2 = & \left(w_1^{(0)}\right)^2 - \left( \beta_1^2 + \beta_2^2 + \gamma_1^2 + \gamma_2^2 \right) \left(w_1^{(0)}\right)^2 - 2 \left( \beta_1\beta_2 - \gamma_1\gamma_2 \right)^2 w_1^{(0)}w_2^{(0)} \, ,
\\
\left|w_2\right|^2 = & \left(w_2^{(0)}\right)^2 - \left( \beta_1^2 + \beta_2^2 + \gamma_1^2 + \gamma_2^2 \right) \left(w_2^{(0)}\right)^2 - 2 \left( \beta_1\beta_2 - \gamma_1\gamma_2 \right)^2 w_1^{(0)}w_2^{(0)} \, ,
\\
w_1w_2+\overline{w_1}\overline{w_2} = & 2w_1^{(0)}w_2^{(0)} - 2 \left( \beta_1^2 + \beta_2^2 + \gamma_1^2 + \gamma_2^2 \right) w_1^{(0)}w_2^{(0)} - 2 \left( \beta_1\beta_2 - \gamma_1\gamma_2 \right) \left[ \left(w_1^{(0)}\right)^2 + \left(w_2^{(0)}\right)^2 \right] \, .
\end{align}
As required, the departures from the fiducial point vanishes at linear order. We can thus write the mass matrix as
\begin{align}
\left. \frac{\partial^2V}{\partial\Psi_i\partial\Psi_i} \right|_0 = & \left[ \frac{\partial{V}}{\partial(w_1+\overline{w_1})} \frac{\partial^2(w_1+\overline{w_1})}{\partial\Psi_i\partial\Psi_j} + \frac{\partial{V}}{\partial\left|w_1\right|^2}
\frac{\partial^2\left|w_1\right|^2}{\partial\Psi_i\partial\Psi_j} \right.
\nonumber\\
& \left.\left. + \frac{\partial{V}}{\partial(w_2+\overline{w_2})} \frac{\partial^2(w_2+\overline{w_2})}{\partial\Psi_i\partial\Psi_j} +
\frac{\partial{V}}{\partial\left|w_2\right|^2} \frac{\partial^2\left|w_2\right|^2}{\partial\Psi_i\partial\Psi_j} +
\frac{\partial{V}}{\partial(w_1w_2+\overline{w_1}\overline{w_2})}
\frac{\partial^2(w_1w_2+\overline{w_1}\overline{w_2})}{\partial\Psi_i\partial\Psi_j} \right] \right|_0 \, ,
\end{align}
where the linear partial derivatives of the brane coordinates do not appear since we have chosen the extremal fiducial trajectory. We can see that in the present angular coordinates $\left\{ \alpha_i,\beta_i,\gamma_i \right\}$, the mass matrix is not diagonal because of the terms which mix different coordinates, such as $(\alpha_1+\alpha_2)^2$, $\beta_1\beta_2$ and $\gamma_1\gamma_2$. Let us rewrite the angular coordinates $\beta_i$ and $\gamma_i$ in terms another set of real coordinates $q_i$ and $r_i$ as
\begin{align}
\beta_{1,2} = & \frac{q_1 \pm r_1}{\sqrt{2}} \, ,
\\
\gamma_{1,2} = & \frac{r_2 \pm q_2}{\sqrt{2}} \, .
\end{align}
Then we immediately find
\begin{align}
\beta_1^2 + \beta_2^2 + \gamma_1^2 + \gamma_2^2 = & q_1^2 + q_2^2 + r_1^2 + r_2^2 \, ,
\\
\beta_1\beta_2 - \gamma_1\gamma_2 = & \frac{q_1^2+q_2^2}{2} - \frac{r_1^2+r_2^2}{2} \, .
\end{align}
We can further define shorthand notations
\begin{align}
\label{DefQ}
q_1^2+q_2^2 \equiv & Q^2 \, ,
\\
\label{DefR}
r_1^2+r_2^2 \equiv & R^2 \, ,
\end{align}
and similarly
\begin{equation}\label{DefP}
\alpha_1+\alpha_2 \equiv P \, ,
\end{equation}
since only these combinations appears. Using these new coordinates $\{P,Q,R\}$, we obtain
\begin{align}
w_1+\overline{w_1} = & 2w_1^{(0)} - \left( P^2+Q^2+R^2 \right)w_1^{(0)} - \left( Q^2-R^2 \right) w_2^{(0)}~,
\\
w_2+\overline{w_2} = & 2w_2^{(0)} - \left( P^2+Q^2+R^2 \right)w_2^{(0)} - \left( Q^2-R^2 \right) w_1^{(0)}~,
\\
\left|w_1\right|^2 = & \left(w_1^{(0)}\right)^2 - \left( Q^2+R^2 \right) \left(w_1^{(0)}\right)^2 - \left( Q^2-R^2 \right) w_1^{(0)}w_2^{(0)} \, ,
\\
\left|w_2\right|^2 = & \left(w_2^{(0)}\right)^2 - \left( Q^2+R^2 \right) \left(w_2^{(0)}\right)^2 - \left( Q^2-R^2 \right) w_1^{(0)}w_2^{(0)} \, ,
\\
w_1w_2+\overline{w_1}\overline{w_2} = & 2w_1^{(0)}w_2^{(0)} - 2 \left( Q^2+R^2 \right) w_1^{(0)}w_2^{(0)} -
\left( Q^2-R^2 \right) \left[ \left(w_1^{(0)}\right)^2 + \left(w_2^{(0)}\right)^2 \right] \, .
\end{align}
These new coordinates therefore make the mass matrix diagonal, with rows and the columns corresponding to
$\{P\,,q_1\,,q_2\,,r_1\,, r_2\}$, as
\begin{equation}\label{mass_matrix}
\left. \frac{\partial^2V}{\partial\Psi_i\partial\Psi_i} \right|_0 = \left.\left(%
\begin{array}{ccccc}
  X_P & 0 & 0 & 0 & 0 \\
  0 & X_Q & 0 & 0 & 0 \\
  0 & 0 & X_Q & 0 & 0\\
  0 & 0 & 0 & X_R & 0\\
  0 & 0 & 0 & 0 & X_R
\end{array}%
\right)\right|
\begin{array}{ccccc}
P\\
q_1\\
q_2\\
r_1\\
r_2
\end{array}
\end{equation}
where
\begin{align}
X_P \equiv & -2 \left[ w_1^{(0)} \left.
\frac{\partial{V}}{\partial(w_1+\overline{w_1})} \right|_0 + w_2^{(0)} \left.
\frac{\partial{V}}{\partial(w_2+\overline{w_2})} \right|_0 \right] \, ,
\\
X_Q \equiv & -2 \left( w_1^{(0)} + w_2^{(0)} \right) \left[ \left.
\frac{\partial{V}}{\partial(w_1+\overline{w_1})} \right|_0 + \left.
\frac{\partial{V}}{\partial(w_2+\overline{w_2})} \right|_0 \right]  + 2\epsilon^2\, \left(
1
 -\cosh{\tau} \right) \left.
\frac{\partial{V}}{\partial(w_1w_2+\overline{w_1}\overline{w_2})} \right|_0
\nonumber\\
& + \left. \frac{\partial{V}}{\partial\left|w_1\right|^2} \right|_0 \left[
-2\left(w_1^{(0)}\right)^2 + \epsilon^2 \right] + \left.
\frac{\partial{V}}{\partial\left|w_2\right|^2} \right|_0 \left[
-2\left(w_2^{(0)}\right)^2 + \epsilon^2 \right] \, ,
\\
X_R \equiv & -2 \left( w_1^{(0)} - w_2^{(0)} \right) \left[ \left.
\frac{\partial{V}}{\partial(w_1+\overline{w_1})} \right|_0  - \left.
\frac{\partial{V}}{\partial(w_2+\overline{w_2})} \right|_0 \right] + 2
\epsilon^2 \left(1 +\cosh{\tau}  \right) \left.
\frac{\partial{V}}{\partial(w_1w_2+\overline{w_1}\overline{w_2})} \right|_0
\nonumber\\
& + \left. \frac{\partial{V}}{\partial\left|w_1\right|^2} \right|_0 \left[
-2\left(w_1^{(0)}\right)^2 - \epsilon^2 \right] + \left.
\frac{\partial{V}}{\partial\left|w_2\right|^2} \right|_0 \left[
-2\left(w_2^{(0)}\right)^2 - \epsilon^2 \right] \, .
\end{align}

It is consequently straightforward to substitute the derivatives of the potential, as well as the expressions for $w_i^{(0)}$ in the previous formulae. Explicitly, we can calculate the angular masses associated with them as
\begin{align}
X_P = &
 \begin{cases}
 \left. \mp2\alpha \left( e^{\tau/2}V_{w_1+\overline{w_1}} - e^{-\tau/2}V_{w_2+\overline{w_2}}
 \right) \right|_0 \, , & \text{(non-delta-flat)}
 \\
 \left. \mp2\alpha \left( e^{-\tau/2}V_{w_1+\overline{w_1}} - e^{\tau/2}V_{w_2+\overline{w_2}}
 \right) \right|_0 \, , & \text{(delta-flat)}
 \end{cases}
\\
X_Q = &
 \begin{cases}
 X_P + 2\alpha^2 \left( 2V_{w_1w_2+\overline{w_1}\overline{w_2}} - e^{\tau}V_{|w_1|^2}
 - e^{-\tau}V_{|w_2|^2} \right) \quad \text{(non-delta-flat)}
 \\
 \left. + \left[ 2\alpha^2 \left( V_{|w_1|^2} + V_{|w_2|^2} \right) - 4\alpha^2\cosh\tau
 V_{w_1w_2+\overline{w_1}\overline{w_2}} \mp 2\alpha \left(
 e^{\tau/2}V_{w_2+\overline{w_2}} - e^{-\tau/2}V_{w_1+\overline{w_1}} \right) \right] \right|_0 \, ,
 \\
 X_P + 2\alpha^2 \left( 2V_{w_1w_2+\overline{w_1}\overline{w_2}} - e^{-\tau}V_{|w_1|^2}
 - e^{\tau}V_{|w_2|^2} \right) \quad \text{(delta-flat)}
 \\
 \left. + \left[ 2\alpha^2 \left( V_{|w_1|^2} + V_{|w_2|^2} \right) - 4\alpha^2\cosh\tau
 V_{w_1w_2+\overline{w_1}\overline{w_2}} \mp 2\alpha \left(
 e^{-\tau/2}V_{w_2+\overline{w_2}} - e^{\tau/2}V_{w_1+\overline{w_1}} \right) \right] \right|_0 \, ,
 \end{cases}
\\
X_R = &
 \begin{cases}
 X_P + 2\alpha^2 \left( 2V_{w_1w_2+\overline{w_1}\overline{w_2}} - e^{\tau}V_{|w_1|^2}
 - e^{-\tau}V_{|w_2|^2} \right) \quad \text{(non-delta-flat)}
 \\
 \left. - \left[ 2\alpha^2 \left( V_{|w_1|^2} + V_{|w_2|^2} \right) - 4\alpha^2\cosh\tau
 V_{w_1w_2+\overline{w_1}\overline{w_2}} \mp 2\alpha \left(
 e^{\tau/2}V_{w_2+\overline{w_2}} - e^{-\tau/2}V_{w_1+\overline{w_1}} \right) \right] \right|_0 \, ,
 \\
 X_P + 2\alpha^2 \left( 2V_{w_1w_2+\overline{w_1}\overline{w_2}} - e^{-\tau}V_{|w_1|^2}
 - e^{\tau}V_{|w_2|^2} \right) \quad \text{(delta-flat)}
 \\
 \left. - \left[ 2\alpha^2 \left( V_{|w_1|^2} + V_{|w_2|^2} \right) - 4\alpha^2\cosh\tau
 V_{w_1w_2+\overline{w_1}\overline{w_2}} \mp 2\alpha \left(
 e^{-\tau/2}V_{w_2+\overline{w_2}} - e^{\tau/2}V_{w_1+\overline{w_1}} \right) \right] \right|_0 \, ,
 \end{cases}
\end{align}
where
\begin{align}
V_{w_1+\overline{w_1}} = & -\frac{\calA}{n} \calG_0^{1/n-1} \left[ \calB - 3ae^{a\sigma}\frac{|W_0|}{|A_0|}\calG_0^{-1/(2n)} - \frac{n-1}{n\calG_0} \left( \calK\coth\tau\calC_0 + \calR_\circ\calD_0 \right) - \calK\coth\tau \right] \, ,
\\
V_{w_2+\overline{w_2}} = & \frac{\calA}{n} \calG_0^{1/n-1} \frac{\calK}{\sinh\tau} \, ,
\\
V_{|w_1|^2} = & -V_{w_1+\overline{w_1}} - \frac{\calA}{n} \calG_0^{1/n-1} \coth\tau \left[ \calK + \calR_\times \left( \frac{1}{\calK} - \coth\tau \right) \right] \, ,
\\
V_{|w_2|^2} = & -\frac{\calA}{n} \calG_0^{1/n-1} \calR_\times \left[ 1 + \coth\tau \left( \frac{1}{\calK} - \coth\tau \right) \right] \, ,
\\
V_{w_1w_2+\overline{w_1}\overline{w_2}} = & -\frac{\calA}{n} \calG_0^{1/n-1}\mathrm{csch}\tau \left[ \calK + \calR_\times \left( \frac{1}{\calK} - \coth\tau \right) \right] \, ,
\end{align}
and
\begin{align}
\calK(\tau) \equiv & \frac{k'}{k''} = \frac{3}{4} \frac{\sinh(2\tau)-2\tau}{\sinh^2\tau} \, ,
\\
\calR_\circ(\tau) \equiv & \frac{r^3}{an\mu^2\gamma k''} = 2\alpha^2\cosh\tau\calR_\times(\tau) \, ,
\\
\calR_\times(\tau) \equiv & \frac{1}{an\gamma k''} = \frac{3^{2/3}}{2^{4/3}} \frac{1}{\beta} \frac{[\sinh(2\tau)-2\tau]^{2/3}}{\sinh^2\tau} \, .
\end{align}
These results are the basic tools for studying the stability of inflationary trajectories along the angular directions.

\subsection{Euler-Rodriguez versus Cayley-Kline parametrization of $SU(2)$ group}
\label{app:versus}

We now discuss in more detail the connection between the Euler-Rodriguez and the Cayley-Kline parametrizations of $SU(2)$ group elements. In particular, we are interested in the relation between the matrices $L_e$ and $R_e$, and the matrices $L_c$ and $R_c$. The analysis here also leads to the derivation of the kinetic terms used in the multi-field analysis in the main text.

Recall that $W$ is obtained separately by using two different parametrizations, (\ref{expMatrixcoord}) and (\ref{forforW}). Since, the expression for $W_0$ (\ref{forforW}) depends on which extremal trajectory one chooses, also the relation between $L_e$ and $L_c$ depends on this choice. From the similarity transformation between $W_0$ and $W_s$, we can find that one possible realization of the relation between $L_c$ and $L_e$ is
\begin{equation}\label{relerck}
L_c = \left\{
 \begin{split}
 &L_e\,i \sigma_2 \, , \qquad \text{(non-delta-flat, lower branch)}
 \\
 &L_e\,i \sigma_1 \, , \qquad \text{(non-delta-flat, upper branch)}
 \\
 &L_e\,\openone \, , \qquad\,\, \text{(delta-flat, lower branch)}
 \\
 &L_e\,\sigma_3 \, , \qquad \text{(delta-flat, upper branch)}
 \end{split}\right.
\end{equation}
and precisely the same for $R_c$ and $R_e$. But other realizations work as well. For the remaining part of this appendix, we focus on the lower branch of the non-delta flat trajectory: this is the case we explicitly considered in Section~\ref{sec:numeric}, and the other cases can be treated in an identical way. The relation between the angles $\theta_i$, $\psi_i$ and $\phi_i$ used in the Cayley-Kline representation and $\alpha_i$, $\beta_i$ and $\gamma_i$ in the Euler-Rodriguez representation is obtained from the first equation in (\ref{relerck}), and reads, both for $i=1$ and 2,
\begin{align}
\label{relangl2p1}
\cos\left(\frac{\theta_i}{2}\right) \cos \left( \frac{\psi_i+\phi_i}{2} \right)
\,= & \,-\gamma_i\,\frac{\sin{|\chi_i|}}{|\chi_i|} \, ,
\\
\label{relangl2p2}
\cos\left(\frac{\theta_i}{2}\right) \sin \left( \frac{\psi_i+\phi_i}{2} \right)
\,= &\,-\beta_i\,\cfrac{\sin{|\chi_i|}}{|\chi_i|} \, ,
\\
\label{relangl2p3}
\sin\left(\frac{\theta_i}{2}\right) \cos \left( \frac{\psi_i-\phi_i}{2} \right)
\,= & \,-\cos{|\chi_i|} \, ,
\\
\label{relangl2p4}
\sin\left(\frac{\theta_i}{2}\right) \sin \left( \frac{\psi_i-\phi_i}{2} \right)
\,= & \,\alpha_i\,\cfrac{\sin{|\chi_i|}}{|\chi_i|} \, .
\end{align}
By means of these relations, we see that the extremal trajectory corresponding to $\alpha_i=\beta_i=\gamma_i=0$ is obtained, for example, by choosing  $\theta^{(0)}_i=\pi$, $\psi^{(0)}_i -\phi^{(0)}_i= 2 \pi$.  By expanding the above expressions around this point, we find
\begin{align}\label{relangl2perts}
\delta \alpha_i\,= & \,\frac{1}{2}\,\left(\delta \phi_i-\delta\psi_i\right) \, ,
\\
\delta \beta_i\,= & \,
\frac{\delta\theta_i}{2}\, \sin \left( \frac{\psi^{(0)}_i+\phi^{(0)}_i}{2} \right) \, ,
\\
\delta \gamma_i\,= & \,
\frac{\delta \theta_i}{2}\, \cos \left( \frac{\psi^{(0)}_i+\phi^{(0)}_i}{2} \right) \, .
\end{align}
These expansions allow to obtain the relation between $\{P,Q,R\}$ and $\{\theta_i,\phi_i,\psi_i\}$. Choosing for definiteness $\psi_i^{(0)} = 2 \pi$, $\phi_i^{(0)} = 0$, the identifications (\ref{DefQ}), (\ref{DefR}) and (\ref{DefP}) yield
\begin{align}
P = & \frac{1}{2}\delta(\phi_1+\phi_2-\psi) \, ,
\\
Q = & -\frac{1}{2\sqrt{2}}\delta(\theta_1-\theta_2) \, ,
\\
R = & -\frac{1}{2\sqrt{2}}\delta(\theta_1+\theta_2) \,
\end{align}
It is easy to check that, near this extremal trajectory, expanding up to quadratic order in angular fluctuations, the pull back-metric (\ref{Defwarpfactor}) reduces to
\begin{equation}\label{ExpAngKE}
\left. d\hat{s}^2_6 \right|_{0} = \frac{\epsilon^{4/3}}{2} K(\tau) \left\{ \frac{d\tau^2+4dP^2}{3K^3(\tau)} + 4 \left[ \cosh^2\left(\frac{\tau}{2}\right)dR^2 + \sinh^2\left(\frac{\tau}{2}\right)dQ^2 \right] \right\} \, .
\end{equation}
This result plays an important role in Section~\ref{sec:numeric}. We have checked that the same expression holds for the expansion around all the extremal trajectories.

To conclude, it is interesting to consider in more detail the case $\alpha_1=\alpha_2=0$, that plays an important role in the discussion of Section~\ref{sec:multifield}. In this case, in order to satisfy (\ref{relangl2p4}), we can choose $\psi_i=2\pi+\phi_i$. Then (\ref{relangl2p1}), (\ref{relangl2p2}) and (\ref{relangl2p3}) simplify, and provide
\begin{align}
\label{relangl2perts21}
|\chi_i| \,= & \,\frac{\pi}{2}-
\frac{\theta_i}{2} \, ,
\\
\label{relangl2perts22}
\beta_i\,= & \,|\chi_i| \sin{\psi_i} \, ,
\\
\label{relangl2perts23}
\gamma_i\,= & \,|\chi_i| \cos{\psi_i} \, .
\end{align}
They are valid for arbitrary values of the angles, inside their interval ranges. Since $\theta_i \in [0,\pi]$, we then learn that $|\chi_i|\in [0,\pi/2]$, or equivalently $\chi_i \in [-\pi/2,\pi/2]$. Moreover, comparing (\ref{relangl2perts22}) and (\ref{relangl2perts23}) with (\ref{defxi1}) and (\ref{defxi2}), we can make the identification $\xi=\psi$ in this specific case. This implies that $\xi$ lies in the range $[0,4\pi]$. The potential is periodic in these angular variables, with periods set by the ranges we have just found. However, let us emphasize once again that the identifications (\ref{relangl2perts21}), (\ref{relangl2perts22}) and (\ref{relangl2perts23}) and the relative field range, are only valid starting from the lower branch of the non-delta-flat trajectory. Making other choices in (\ref{relerck}) lead to different identifications. On the other hand, it is simple to work out them following the same procedure as
above.

\section{Stabilized value of $\sigma$}
\label{secstabsigma}

In this Appendix, we analyse the stabilisation of $\sigma$ in the limit $\tau\to 0$. In the absence of uplifting, the stabilized value of $\sigma$ at $\tau = 0$ is given by the condition
\begin{equation}
\left. \frac{\partial V_F(0,\sigma)}{\partial \sigma} \right|_{\sigma = \sigma_F}\,=\,0.
\end{equation}
From (\ref{VFtausigma}), one easily finds
\begin{equation}\label{valssigF}
3ae^{a\sigma_F}\frac{|W_0|}{|A_0|}\calG_0^{-1/(2n)}(0) = 3+2a\sigma_F+\frac{1+a\sigma_F}{2+a\sigma_F}\calF(0) \, .
\end{equation}
Then, we add to $V_F$ an uplifting term of the form
\begin{equation}
V_{D}(\tau,\sigma)\,=\,\frac{D(\tau)}{U^2(\tau,\sigma)} \, ,
\end{equation}
from which we define the uplifting ratio
\begin{equation}\label{Defs}
{s} \equiv \frac{D(0)/U^2(0, \sigma_F)}{|V_{F}(0,\sigma_F)|} \, .
\end{equation}
In order to avoid a runaway decompactification, $s$ should be of order one, $1\le s \le 3$~\cite{Kachru:2003aw}.

The stabilized value of the K\"ahler modulus after uplifting, which we call $\sigma_0 \equiv \sigma_F+\delta \sigma$~\cite{Baumann:2007ah}, is found from the condition
\begin{equation}
\left. \frac{\partial V}{\partial\sigma} \right|_{\sigma_0} \,=\,0 \approx
\left. \frac{\partial^2 V_F}{\partial\sigma^2} \right|_{\sigma_F}\,\delta
\sigma + \left. \frac{\partial V_D}{\partial \sigma} \right|_{\sigma_0} \, ,
\end{equation}
where
\begin{equation}
\left. \frac{\partial V_D}{\partial\sigma} \right|_{\sigma_0}\,\approx\,
-\frac{2 V_D}{\sigma_F}\,\left( 1-3\frac{\delta\sigma}{\sigma_F} \right) \, .
\end{equation}
Then in the limit $\delta\sigma/\sigma_F \ll 1$, we obtain
\begin{equation}\label{fordesig}
\delta \sigma \,=\,\frac{\sigma_F}{3+
\left. \left[\sigma^2 \left(\partial^2V_F/\partial \sigma^2\right)/ 2V_D
\,\right] \right|_{\sigma_F} } \, .
\end{equation}
In the limit of large $a \sigma_F$, we can show that
\begin{equation}
\left. \frac{\partial^2V_F}{\partial \sigma^2} \right|_{\sigma_F} = 2 a^2\, |V_F(0,\sigma_F)| \, .
\end{equation}
Plugging this expression into (\ref{fordesig}) and using (\ref{Defs}), we obtain
\begin{equation}
a \delta \sigma \approx \frac{s}{a\,\sigma_F} \, .
\end{equation}
Then it is then easy to see that $\delta \sigma\ll 1$ induces the following contribution to the stabilized modulus,
\begin{equation}
3ae^{a\sigma_0}\frac{|W_0|}{|A_0|}\calG_0^{-1/(2n)}(0) \approx 3 + 2a\sigma_0 + \frac{1+a\sigma_F}{2+a\sigma_F}\calF(0) + 2s \, .
\label{stabilisation}
\end{equation}
The stabilized value of $\sigma$, in the limit of $\tau\to0$, is given by the solution of (\ref{stabilisation}).

The dependence of $\sigma$ on $\tau$ is more difficult to determine, for non-zero $\tau$, and a numerical analysis is needed. See however Refs.~\cite{Chen:2008ada,Chen:2008ai} for some analytical results.

\end{document}